\documentclass[nofootinbib,aps,prl,twocolumn,superscriptaddress,longbibliography]{revtex4-2}
\usepackage{graphicx}  % needed for figures
\usepackage{subfigure}
\usepackage{dcolumn}   % needed for some tables
\usepackage{bm}        % for math
\usepackage{amssymb}   % for math
\usepackage{tabularx,booktabs,ragged2e,multirow}  
\newcolumntype{C}{>{\centering\arraybackslash}X}
\usepackage{multirow}
\usepackage{comment}
\usepackage[pdfencoding=auto, psdextra]{hyperref}
\hypersetup{
  colorlinks   = true, %Colours links instead of ugly boxes
  urlcolor     = blue, %Colour for external hyperlinks
  linkcolor    = orange, %Colour of internal links
  citecolor   = cyan %Colour of citations
}
\usepackage[normalem]{ulem}
\usepackage{color}
\usepackage{xcolor}
\usepackage{amsmath}
\usepackage{mathrsfs}
\usepackage{mathcomp}
\usepackage{mathdots}
\usepackage{siunitx}
\usepackage{textcomp}
\usepackage{dsfont}
\usepackage{esint}
\usepackage{braket}
\usepackage{textgreek}
\usepackage{lipsum}
\usepackage{centernot}
\usepackage{cancel}
\DeclareMathOperator{\sgn}{sgn}

\begin{document}
\newcommand{\br}{{\bm r}}
\newcommand{\bR}{{\bm R}}
\newcommand{\bk}{{\bm k}}
\newcommand{\bq}{{\bm q}}
\newcommand{\bp}{{\bm p}}
\newcommand{\bv}{{\bm v}}
\newcommand{\bmm}{{\bm m}}
\newcommand{\bA}{{\bm A}}
\newcommand{\bE}{{\bm E}}
\newcommand{\bB}{{\bm B}}
\newcommand{\bH}{{\bm H}}
\newcommand{\bd}{{\bm d}}
\newcommand{\bw}{{\bm w}}
\newcommand{\bu}{{\bm u}}
\newcommand{\brho}{{\bm \rho}}
\newcommand{\bzero}{{\bm 0}}
\newcommand{\bOmega}{{\bm \Omega}}
\newcommand{\bsigma}{{\bm \sigma}}
\newcommand{\bJ}{{\bm J}}
\newcommand{\bL}{{\bm L}}
\newcommand{\bS}{{\bm S}}
\newcommand{\bP}{{\bm P}}
\newcommand{\bZ}{{\bm Z}}
\newcommand\dd{\mathrm{d}}
\newcommand\ii{\mathrm{i}}
\newcommand\ee{\mathrm{e}}
\newcommand\zz{\mathtt{z}}

%\preprint{}

%\title{Optically generating spiral structure in pumped topological insulator surface states with spin- and angle-resolved photoemission spectroscopy}
%\title{Nonequilibrium tornadoes in optically pumped magnetic topological insulator surface states with spin- and angle-resolved photoemission spectroscopy}
\title{Nonequilibrium topological spin textures in momentum space}

\author{Xiao-Xiao Zhang}
%\email{xiaoxiao.zhang@ubc.ca}
%\affiliation{Department of Physics and Astronomy \& Stewart Blusson Quantum Matter Institute, University of British Columbia, Vancouver, BC, V6T 1Z4 Canada}
\affiliation{RIKEN Center for Emergent Matter Science (CEMS), 2-1 Hirosawa, Wako, Saitama 351-0198, Japan}
%\affiliation{Department of Applied Physics, The University of Tokyo, 7-3-1 Hongo, Bunkyo-ku, Tokyo 113-8656, Japan}
\author{Naoto Nagaosa}
%\email{nagaosa@ap.t.u-tokyo.ac.jp}
\affiliation{Department of Applied Physics, The University of Tokyo, 7-3-1 Hongo, Bunkyo-ku, Tokyo 113-8656, Japan}
\affiliation{RIKEN Center for Emergent Matter Science (CEMS), 2-1 Hirosawa, Wako, Saitama 351-0198, Japan}
%\date{\today}
\makeatletter
\let\newtitle\@title
\let\newauthor\@author
\def\ExtendSymbol#1#2#3#4#5{\ext@arrow 0099{\arrowfill@#1#2#3}{#4}{#5}}
\newcommand\LongEqual[2][]{\ExtendSymbol{=}{=}{=}{#1}{#2}}
\newcommand\LongArrow[2][]{\ExtendSymbol{-}{-}{\rightarrow}{#1}{#2}}
\newcommand{\cev}[1]{\reflectbox{\ensuremath{\vec{\reflectbox{\ensuremath{#1}}}}}}
\newcommand{\red}[1]{\textcolor{red}{#1}} %for displaying red texts
\newcommand{\green}[1]{\textcolor{green}{#1}} %for displaying green texts
\newcommand{\blue}[1]{\textcolor{blue}{#1}} %for displaying blue texts
\newcommand{\mytitle}[1]{\textcolor{orange}{\textit{#1}}}
\newcommand{\mycomment}[1]{} %for commenting out
\newcommand{\note}[1]{ \textbf{\color{blue}#1}}
\newcommand{\warn}[1]{ \textbf{\color{red}#1}}
\newcommand{\tensorhead}[1]{\overset{\text{\tiny$\leftrightarrow$}}{#1}}
\makeatother

\begin{abstract}
%The surface state of a genuine or magnetically doped topological insulator is a prototypical low-dimensional quantum system readily accessible from various surface spectroscopic techniques. We theoretically propose that contemporary optical pump-probe-type experiment can extend the system to a capable platform for controlled light-matter interaction and photoinduced ultrafast quantum coherent nonequilibrium dynamics, for which we exemplify with the time-resolved angle-resolved photoemission spectroscopy (ARPES) in the spin channel with linear and circular polarization of the laser injection.  Assuming optical long-wavelength limit appropriate to the common ARPES laser beam, calculation shows a surprisingly rich tornado-like evolving pattern in both the normal and in-plane spin-resolved ARPES signal. In the Keldysh formalism, we systematically reveal its origin as a unique nonequilibrium photoinduced topological winding phenomenon in the momentum space. In addition to a remarkable nonequilibrium tomography of all the intrinsic and extrinsic helicities of the material and light, the in-plane signal intriguingly exhibits a dichroic $\mathbb{Z}_2$-like topological winding response.

Nonequilibrium quantum dynamics of many-body systems is the frontier of condensed matter physics; recent advances in various time-resolved spectroscopic techniques continue to reveal rich phenomena. Angle-resolved photoemission spectroscopy (ARPES) as one powerful technique can resolve electronic energy, momentum, and spin along the time axis after excitation. However, dynamics of spin textures in momentum space remains mostly unexplored. Here we demonstrate theoretically that the photoexcited surface state of genuine or magnetically doped topological insulators shows novel topological spin textures, i.e., tornado-like patterns, in the spin-resolved ARPES. We systematically reveal its origin as a unique nonequilibrium photoinduced topological winding phenomenon. 
As all intrinsic and extrinsic topological helicity factors of both material and light are embedded in a robust and delicate manner, the tornado patterns not only allow a remarkable tomography of these important system information, but also enable various unique dichroic topological switchings of the momentum-space spin texture.
%Besides a remarkable tomography of all intrinsic and extrinsic topological helicity factors of material and light from the tornado patterns, the in-plane signal dichroically switches between trivial and topological responses as an unusual topological optical activity. 
These results open a new direction of nonequilibrium topological spin states in quantum materials.

\end{abstract}
%\pacs{71.10.Pm, 71.27.+a, 72.15.Nj, 72.15.Rn}
\keywords{}

\maketitle
%\tableofcontents
\let\oldaddcontentsline\addcontentsline% Store \addcontentsline
\renewcommand{\addcontentsline}[3]{}% Make \addcontentsline a no-op

%\mytitle{Introduction}.--
\section*{Introduction}
The recent %\cite{Freericks2009,Freericks2015,Freericks2016,Reimann2018,Jozwiak2016,Cacho2015,Stefanucci2015,Bar_Gill_2013,Sigillito_2015,Hosur2011,Iyer2018,liu_2013} 
decade has witnessed significant advances in the detection means of ultrafast light-induced phenomena\cite{Giannetti2016,Nicoletti2016\mycomment{,Udina2019}} in terms of time-resolved spectroscopic techniques including angle-resolved photoemission spectroscopy (ARPES)\cite{Smallwood2012,Gedik2017,Sobota2021}, %optical bulk reflection or thin-film transmission measurements, 
%electron diffuse scattering\cite{Stern2018,Cotret2019}, 
terahertz pump-probe scanning-tunneling microscopy and optical conductivity measurement\cite{Loth2010,Cocker2013,Eisele2014,Mitrano2016}, %resonant inelastic X-ray scattering\cite{Cao2019,Mitrano2020}, 
etc. 
Unprecedented precise access into the inherently time-dependent phenomena is beneficial and important to both the fundamental interest in nonequilibrium physics and the practical connection to ultrafast manipulation of novel quantum degrees of freedom towards %next-generation information technology
application\cite{Ostroverkhova2016,Basov2017,Tokura2017}. To this end, a robust low-dimensional nontrivial system would be a versatile playground for such surface-sensitive pump-probe-type investigation. The protected surface state of topological insulator fits into this role for its long enough mean free path and lifetime and also for excluding the insulating and spin-degenerate bulk influence\cite{Topo1,Topo2,Neupane2015}. Tunable exchange gap from controlling magnetic doping further allows for demonstrating both massless and massive Dirac physics\cite{Chen2010,Chang2013,Tokura2019,Wang2021\mycomment{,DiracFermion1,DiracFermion2}}. 

However, nonequilibrium spin dynamics is usually studied in time domain or real space only\cite{Kirilyuk2010,Kirilyuk2013}.
For the surface state, it has been focused on the equilibrium spin-orbit coupling features\cite{Hsieh2009,Xu2016} and the photodriven steady-state or highly pumped charge current responses\cite{Gedik,Oka2019,Hosur2011,McIver2011,Reimann2018,Yu2019a}. The nonequilibrium phenomena of light-matter interaction in this system remain largely buried partially due to the little appreciated spin-channel physics. 
In fact, such information connects well to the state-of-art experimental reach, e.g., spin-resolved ARPES (SARPES) has been established in equilibrium and as well extended to time-dependent measurement well below picosecond resolution\cite{Hsieh2009,Xu2016,Schuler_2020,Cacho2015,Zhang2018,Jozwiak2016,Nie2019,Sobota2021}.
As an example of the new front of nonequilibrium quantum dynamics of topological matters, we draw attention to this highly informative time-dependent signal in an optical pump-probe experiment upon the surface state.

In particular, we simulate the irradiation of a terahertz short laser pulse, which can be either linearly polarized (LP) or circularly polarized (CP)\cite{Lv2019}, to pump across the exchange gap; then detect the SARPES signal after a controllable delay time with a probe pulse. Apart from possible resonant transition, virtual excitation at the early stage of time evolution is a purely quantum mechanical effect and can turn the system into a many-particle coherent nonequilibrium state.
Surprisingly, the SARPES signal exhibits robust and topological tornado-like spiral structures in the two-dimensional (2D) momentum $\bk$-space, which can be characterized by topological indices. This happens in both the normal and in-plane spin channels and embeds a delicate relation to three helicity factors determining the pumped system: intrinsic helicity of the surface state $\chi=\pm1$, sign of the Dirac mass $\nu=\sgn{(m)}$, and extrinsic helicity $\tau=0,\pm1$ respectively for LP and right or left CP lights. %They are encoded in both the radial $k$ and azimuthal $\theta$ directions of $\bk$-space. 
Depending on these, the novel tornado-like responses can dichotomously change characteristic winding senses and even dichroically switch between topological and trivial as a $\mathbb{Z}_2$-like topological optical activity.

\begin{figure*}[hbt]
\includegraphics[width=15cm]{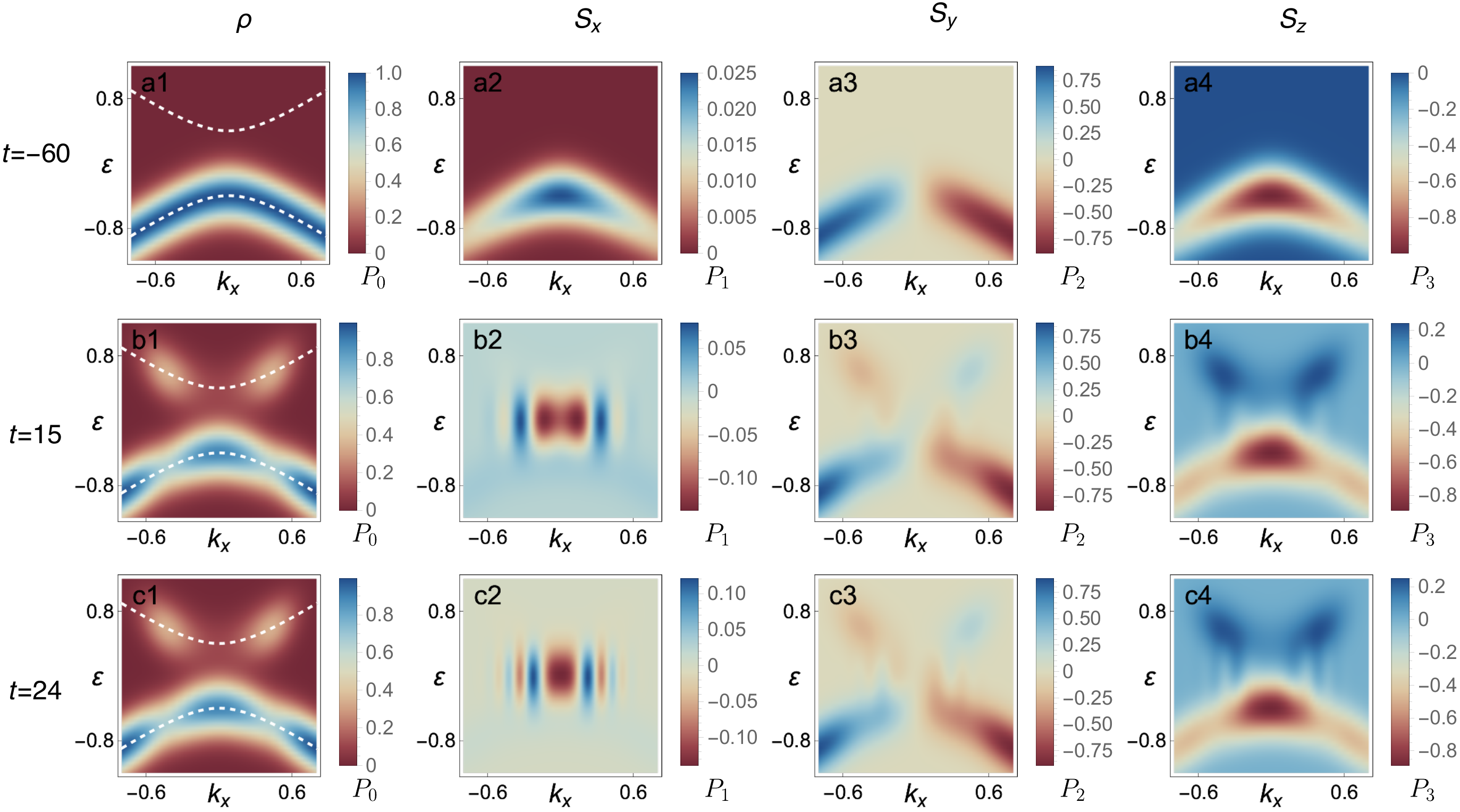}
\caption{\textbf{Nonequilibrium spin-resolved ARPES (SARPES) signals in the $(\varepsilon,k_x)$-plane.} $P_0,P_1,P_2,P_3$ 
successively in the density $\rho$ and spin $\bS$ channels of a magnetic topological insulator surface state at three different times. White dashed curves in panels (a1,b1,c1) indicate the surface state band dispersion. The band broadening originates from finite probe pulse width. 
Parameters are $\chi=\tau=1,t_0=t_\mathrm{pb}=3,\Omega=1.2,v=1,m=0.4,A_0=0.1,k_y=0.01,\beta=50,\mu=0,e=\hbar=k_B=1$.
(a) $t=-60$ signal before pump pulse irradiation exhibits equilibrium response: only lower band is visible due to relatively low temperature specified by $\beta=1/(k_BT)$ and in-gap chemical potential. The $90$-degree out-of-phase spin-momentum locking manifests in the spin channels: $P_1$ is weak compared to others due to small $k_y$; %close to zero where $P_1$ will vanish; 
$P_2$ reverses sign between positive and negative $k_x$-axis; $P_3$ is made finite purely by the finite exchange gap. (b) At $t=15$ after the pump pulse centered at $t=0$ almost fully decays, resonant real transition appears as two spots in the upper band in $P_0$. The spin channels exhibit a signal hot region centered at $\varepsilon=0$ and $k_x=0$, which is oscillatory in time and momentum. This is clearly seen in $P_1$ for the weak background from real band occupations, compared to $P_2,P_3$. (c) At a later time $t=24$, while the density channel remains nearly time-independent after the pumping process, the hot region signals in the spin channel evolve in time with increasing fine structures, implying that it originates mainly from virtual excitations and the coherent quantum oscillation correlated in momentum space.
}\label{Fig:Ekx_main}
\end{figure*}

\section*{Results}
%\mytitle{Model and time evolution}.--
\subsection*{Model and time evolution}
We consider the 2D massive Dirac model and henceforth set $\hbar=1$
\begin{equation}
    H_0(\bk)=\bd(\bk)\cdot\bsigma=v(k_x\sigma_2 - \chi k_y\sigma_1) + m\sigma_3
\end{equation}
to represent the surface state with spin Pauli matrices $(\sigma_0,\bsigma)=(I,\sigma_1,\sigma_2,\sigma_3)$. We include the $\chi=-1$ case possible when $C_{n>2}$ rotational symmetries are broken. The two bands $\varepsilon_{\bk\pm} =d_0(k)\pm d(\bk)\mycomment{=\pm\sqrt{v^2k^2+m^2}}$ if we include the spin-independent quadratic term $d_0(k)\sigma_0$, which is henceforth dropped as it does not affect spin channel response from interband transitions. The hexagonal warping strength $c_6$ measured in the dimensionless quantity $c_6k_0^2/v \ll 1$ makes it negligible with the characteristic wavenumber $k_0$ introduced later\cite{Fu2009,Liu2010}. Therefore, our prediction is fully based on the leading order response in real systems. 
The ARPES light source typically bears a beam spot size $10\textrm{--}100\mathrm{\mu m}$ upon the sample\cite{Giannetti2016,Lv2019,Cattelan2018,Sobota2021}, which requests one to consider physical phenomena at the optical long-wavelength limit as the experimentally most relevant scenario, in contrast to the otherwise interesting space-resolved nano-ARPES or scanning Kerr magnetooptic microscopy study\cite{Keatley2017,McCormick2017,Yamamoto2018}.
%According to the foregoing discussion when optical wavelength $\lambda\gg 2\pi/k_0$, 
We thus introduce a spatially uniform Gaussian vector potential for the pump pulse vertically shone onto the $xy$-plane
%\begin{equation}\label{op}
$\bA(t) = A_0\exp(-t^2/2t_0^2)\left[\hat{x}\cos\Omega t + \tau\,\hat{y}\sin\Omega t\right]$, 
%\end{equation}
where $\tau=0,\pm1$ %respectively corresponds to $x$-polarized LP light and $\mathrm{CP}^\mathrm{R/L}$ light 
and $t_0$ the temporary width. The conserved momentum enables us to derive the full electromagnetically coupled Hamiltonian from Peierls substitution
\begin{equation}\label{eq:H_main}
	%H_{e-op} = \frac{e}{\hbar}\psi^{\dagger}(\bk)(\partial_{k_\mu} H_0)_{\frac{\bk+\bk'}{2}}\psi(\bk^{\prime})A_{\mu}(\bk-\bk^{\prime})
	\hat{H}(t) = \sum_\bk\psi^{\dagger}(\bk) \left[ H_0(\bk) + \mycomment{\frac{e}{\hbar}}e \partial_{\bk} H_0(\bk) \cdot \bA(t) \right] \psi(\bk)
\end{equation}
with $\psi(\bk)=(\psi_{\bk\uparrow},\psi_{\bk\downarrow})^\mathrm{T}$. 
The time-dependent spinor operator $\psi_{\bk\alpha}(t)$ for $\alpha=\uparrow/\downarrow$ can be obtained via the equation of motion, which relates to the double-time matrix removal Green's function with nonequilibrium occupation and excitation information $G_{\alpha\beta}^<(\bk,t_1,t_2)=\ii \langle \psi_{\bk\beta}^\dag(t_2) \psi_{\bk\alpha}(t_1) \rangle$\cite{Kamenev2009,Stefanucci2015} (see Methods).

%\mytitle{Time-dependent SARPES signal}.--
\subsection*{Time-dependent SARPES signal}

We generalize the time-resolved ARPES theory to obtain the time-dependent SARPES intensity matrix\cite{Freericks2009,Kemper2018} $P(\varepsilon, \bk,t) = -\mycomment{\frac{\ii}{\hbar^2}}\ii\int\mycomment{_{-\infty}^{\infty}} \mycomment{\int\mycomment{_{-\infty}^{\infty}}} \dd t_1 \dd t_2\, \ee^{\ii\varepsilon (t_1-t_2)} s(t_1-t)s(t_2-t) G^<(\bk, t_1, t_2)$ with $s(t)=(2\pi t_\mathrm{pb}^2)^{-\frac{1}{2}}\ee^{-t^2/2 t_\mathrm{pb}^2}$ the isotropic probe pulse of width $t_\mathrm{pb}$ and the spin-polarized photocurrent intensity $I_\alpha \propto P_{\alpha\alpha}$ (see \ref{SM:SARPES}). Then we define 
\begin{equation}\label{eq:I_photo1_main}
    P_i(\varepsilon, \bk,t) = \,\mathrm{Tr}[\sigma_i P(\varepsilon, \bk,t)], \qquad i=0,1,2,3
\end{equation}
successively for the density and three spin channels
to be our main focus since the SARPES polarization reads, e.g., for $z$-direction, $\mathrm{P}_z=\frac{I_\uparrow-I_\downarrow}{I_\uparrow+I_\downarrow} = \frac{P_3}{P_0}$.
As we mainly consider a probe pulse well separated from the pump pulse ($t\gg t_0$), 
we can stick to the present Hamiltonian gauge and are free from gauge invariance issue\cite{Freericks2015,Freericks2016}.
%we evade the gauge-invariant ansatz that does not guarantee $I_\alpha$'s physical positivity for multiband cases\cite{Freericks2015,Freericks2016}. 

The pump field renders the original Dirac bands no longer eigenstates and occupation can in general change: in the $(\varepsilon,\bk)$-hyperplane, not only on-resonance real transition can happen when the gap $\Delta=2m<\Omega$, which is the case shown in Fig.~\ref{Fig:Ekx_main}, but also off-resonance virtual excitations significantly contribute, constituting a transient redistribution along the $\varepsilon$-axis per the particle conservation as a sum-rule-like constraint.
%In addition, the probe pulse finite width $t_\mathrm{pb}$, which relaxes energy conservation, broadens this transient process and relevant energy levels in detection and thus inverse-proportionally controls the virtual excitation amount in particular. 
After the pump field fully decays, Dirac bands return to be eigenstates. For the density channel, shown in Figs.~\ref{Fig:Ekx_main}(a1,b1,c1), this implies that, except for resonant interband transition, the signal should mostly become stable %return to its initial value 
after the pumping transients. %of being empty, filled, or off-shell, 
%up to probe smearing.
However, in the spin channel pumping has already left relics of light-matter interaction. Each momentum accommodates a two-level system and is subject to the common photoexcitation. This leads to a highly nontrivial correlation of excited spin-orbit-coupled states in $\bk$-space as the central cause of the SARPES tornado textures discussed below. 
Indeed, collective quantum oscillation effect can emerge near some hot region in the $(\varepsilon,\bk)$-hyperplane of SARPES, centered at the band midpoint as shown in Figs.~\ref{Fig:Ekx_main}(b2-4,c2-4). This is because the spin channel extracts the Rabi-like oscillatory information %instead of the band-basis trivial evolution as $\hat{H}$ loses time-dependence after the pump pulse. 
due to interband coherence even as $\hat{H}$ loses time-dependence after the pump pulse.
%for the time-independent $\hat{H}$ after the pump pulse.
Note also that, as is physically originated from the spin-channel interband quantum oscillation, the real resonant pumping, if any, is insignificant for the hot region signals, which will also become clear later with the analytical result \eqref{eq:P_TI_main}.

The probe pulse width $t_\mathrm{pb}$ is a double-edged sword per the uncertainty relation: smaller $t_\mathrm{pb}$ gives better time resolution but less energy resolution and vice versa. 
It thus broadens the transient process and smears the SARPES energy levels. %in detection and thus inverse-proportionally controls the virtual excitation amount in particular
%In general, we would need a narrow enough probe pulse in order to pick up the quantum oscillations typically with period $T_0=2\pi/\Delta$, e.g., $t_\mathrm{pb}\sim T_0/3$. 
Futhermore, certain amount of relaxed energy conservation $\delta\varepsilon\sim 2\pi/t_\mathrm{pb}$ and the associated momentum range $\delta k\propto\delta\varepsilon/v$ can actually enhance the signal from off-resonance oscillations and provide the hot region characteristic scales, because energy-sharp bands are %more precisely in the band basis 
incapable of capturing the quantum oscillations.
%a more complex momentum range $\delta k\propto\delta\varepsilon/v$ for fixed $t_\mathrm{pb}/t_0$, which provide the hot region characteristic scales in $(\varepsilon,\bk)$-hyperplane,
%are actually beneficial to enhancing the signal from off-resonance oscillations as aforementioned, because energy-sharp SARPES bands are more precisely in the band basis incapable of appropriately capturing the quantum oscillation in the spin channel. 
Certainly, too poor energy resolution would otherwise mix contributions, for instance, from both the lower band and the possible higher occupation due to resonant transition.  %$\tau_\mathrm{prb}=\tau_\mathrm{pmp}=3$ when gap $\Delta=0.6$, i.e., typical period $T_0 \lesssim 10$
We also emphasize that this quantum nonequilibrium phenomenon goes beyond the semiclassical picture\cite{Xiao2010}: neither the pumping process nor the interband coherent dynamics at any time can be captured by the wavepacket description within a single band. Direct evidence is the anomalous tornado rotation as quasiparticle trajectory, %with a fixed spin
which is otherwise not expected after the driving electric field in the pump pulse dies out.

\begin{figure*}[hbt]
\includegraphics[width=17.9cm]{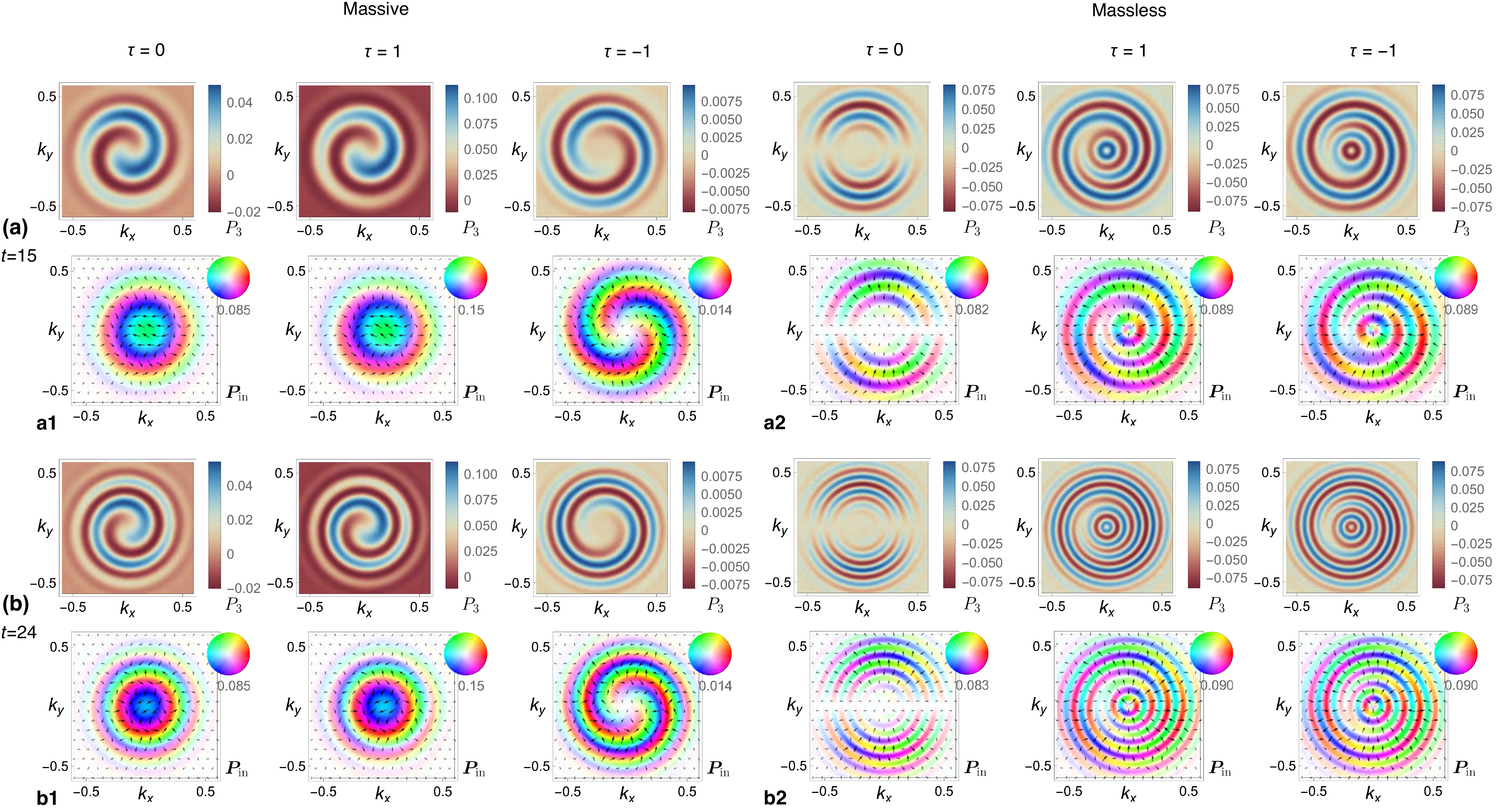}
\caption{\textbf{Nonequilibrium tornado-like responses in the $(k_x,k_y)$-plane.} Equilibrium response subtracted SARPES signals (normal-direction $P_3$ and in-plane $\bP_\mathrm{in}=(P_1,P_2)$) at (a) $t=15$ (b) $t=24$ after the pump pulse. Energy cut at band midpoint $\varepsilon=0$ is adopted without loss of generality. (a1,b1) Positive mass ($\nu=1$) and (a2,b2) massless case for fixed surface state helicity $\chi=1$. Pump light dependence ($\tau=0,\pm1$ for LP along $x$-axis and right/left CP) displayed across the columns.
Scalar $P_3$ plotted for spin-$S_z$ signal. In-plane spin orientation angle $\phi=\tan^{-1}{\bP_\mathrm{in}}\mycomment{\in(-\pi,\pi]}$ plotted according to the rainbow color wheel inset; magnitude $|\bP_\mathrm{in}|$ shown in opacity with maximal $|\bP_\mathrm{in}|$ indicated below each color wheel. Selected $\bP_\mathrm{in}$ vector arrows are shown with corresponding magnitude and orientation. See Fig.~\ref{Fig:tornado_illustrate_main}(d) for enlarged illustration. Topological tornado-like spirals appear except the gapless case under LP light. As time elapses, from (a) to (b), tornadoes evolve and rotate and more tornado arms will be accommodated within a fixed $\bk$-space region. Tornado responses as the distinguishing feature in relation to all three helicity factors are summarized in Table.~\ref{Table:feature_summary}. Dichroic $P_3$-tornado switching helicity with different CP lights [(a1,b1) $\tau=\pm1$ case of $P_3$] is in stark contrast to the $\mathbb{Z}_2$-like $\bP_\mathrm{in}$-tornado, which appears only under one particular CP light in the gapful case [(a1,b1) $\tau=-1$ case of $\bP_\mathrm{in}$]. $\phi$ in the gapless case exhibits $\pi$-jump, due to vanishing $\bP_\mathrm{in}$, along the radial direction once it goes across a spiral arm [(a2,b2) case of $\bP_\mathrm{in}$].
Parameters same as Fig.~\ref{Fig:Ekx_main}.}\label{Fig:tornado_main}
\end{figure*}

\subsection*{Nonequilibrium tornado responses}

\begin{figure}[hbt]
\includegraphics[width=8.7cm]{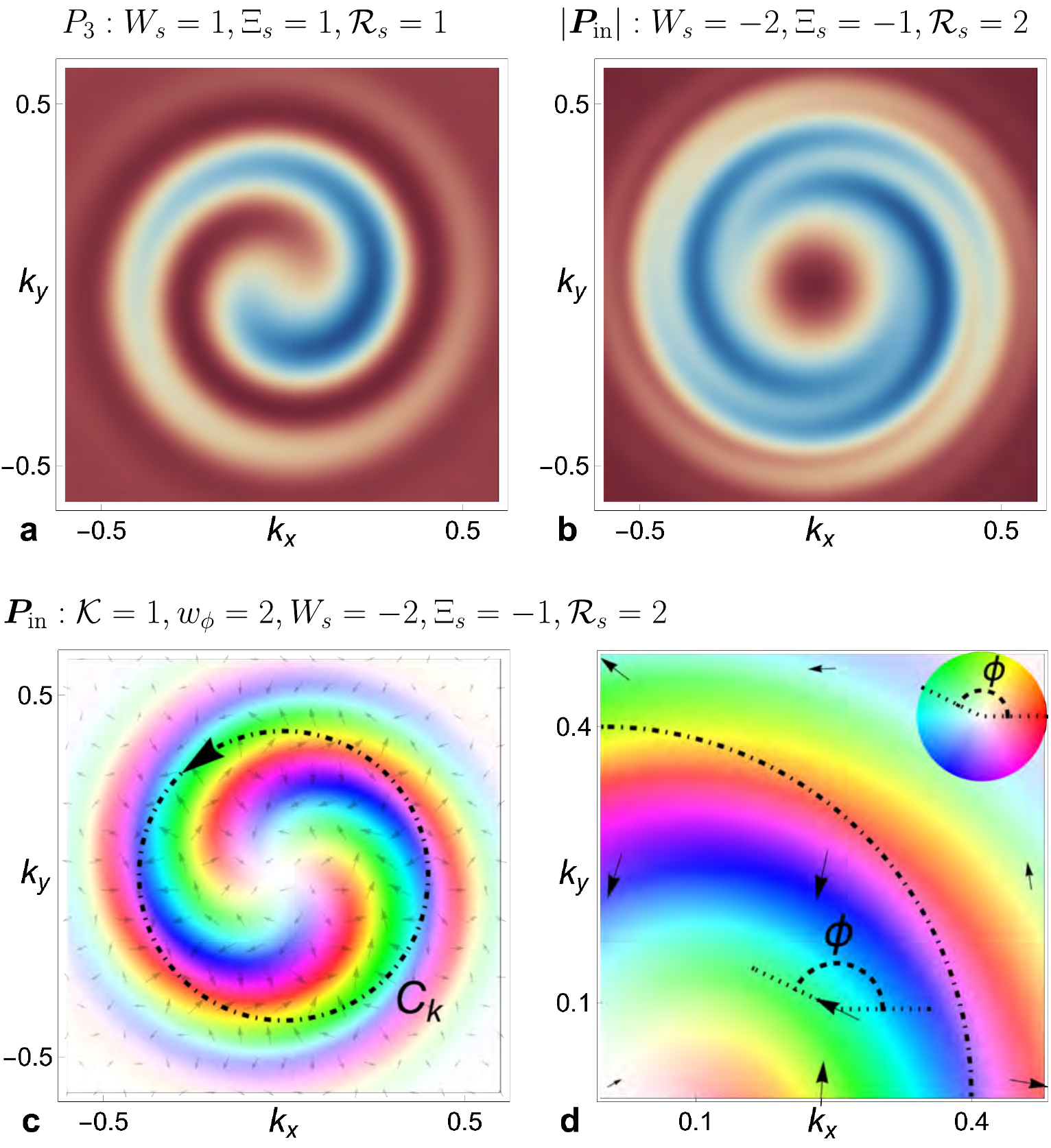}
\caption{\textbf{Topological tornado indices illustrated in representative cases.} Parameters same as Fig.~\ref{Fig:tornado_main}(a1) massive case at $t=15$. Scale legends are omitted for simplicity as they are unimportant for the robust tornado features.
The spiral winding $W_s$ is common for scalar signal (a) $P_3$ for $\tau=1$ or (b) $|\bP_\mathrm{in}|$ for $\tau=-1$ and vectorial in-plane signal (c) $\bP_\mathrm{in}$ for $\tau=-1$. $W_s$ determines the tornado spiral helicity $\Xi_s=\sgn{W_s}$ and the number $\mathcal{R}_s=|W_s|$ of repeating spiral arms. For the vectorial signal, more specific radial ordering $\mathcal{K}$ and azimuthal winding $w_\phi$ also exist and are combined to give $W_s$. (c) shows the counterclockwise circle $C_k$ used in defining winding numbers. (d) zooms in the top right quadrant of (c) and exemplifies a particular vector $\bP_\mathrm{in}$ and its orientation angle $\phi$ together with the rainbow color wheel inset. 
%However, the rotation sense of tornado arm in $\phi$ always follows that of $|\bP_\mathrm{in}|$, which is a result of the locking condition between $\chi,\nu,\tau$ of showing topological $\bP_\mathrm{in}$. In fact, such tornado arm akin to the contour of $\phi$ is merely determined by the combination of its radial variation ($\nu$) and azimuthal variation ($\chi$).
}\label{Fig:tornado_illustrate_main}
\end{figure}

The most interesting information lies in the $\bk$-space spin texture $\bP(\varepsilon,\bk,t)=(P_1,P_2,P_3)$ within an energy slice in the hot region, where robust tornado-like structures widely appears as shown in Fig.~\ref{Fig:tornado_main} (see \ref{Fig:tornado_main_rvs_chi} \ref{Fig:tornado_main_rvs_nu} \ref{Fig:tornado_main_rvs_chi_nu} for cases with different $\chi,\nu$). 
Such energy-momentum hot region lies in general away from where resonant real transitions happen since the tornado mainly originates from coherent virtual excitations, which will be seen also from analytical results.
As aforementioned, there are three helicity factors $\chi,\nu,\tau$ at play during the light-matter interaction, for which the subsequent nonequilibrium tornado response turns out to be an exceptionally apt and reliable bookkeeper. 
For any tornado pattern, one can intuitively identify the rotation sense helicity $\Xi_s=\pm1$ of the spiral and the number $\mathcal{R}_s$ of repeating spiral arms. Practically, $\Xi_s=\sgn[\partial k^*/\partial\theta_\bk]$ with $\theta_\bk$ the azimuthal angle of $\bk$ and $k^*(\theta_\bk)$ any polar-coordinate contour line in a spiral arm. These two lead to the universal topological spiral winding number 
\begin{equation}\label{eq:spiral_winding}
    W_s=\Xi_s\mathcal{R}_s.
\end{equation}
We exemplify these quantities in Fig.~\ref{Fig:tornado_illustrate_main}.
For the in-plane orientation $\phi(\bk)=\tan^{-1}\bP_\mathrm{in}(\bk)$ of the vector field $\bP_\mathrm{in}=(P_1,P_2)$, $W_s$ is readily determined by a combination of $\phi$'s radial and azimuthal variation. $\phi(\bk)$ has a definite ordering, $\mathcal{K}=\sgn(\partial_k \phi)$, i.e., the rainbow order along the radius in our illustration. The latter is encoded in a topological circular winding number 
\begin{equation}\label{eq:tornado_winding}
    w_\phi=\frac{1}{2\pi}\int_{C_k} \dd\bk\cdot\nabla \phi(\bk)
\end{equation}
along a counterclockwise circle $C_k$ of any radius $k$ in the 2D $\bk$-plane. We hence obtain $W_s=-\mathcal{K}w_\phi$. Note that, depending on the helicity factors, any two of $\mathcal{K},w_\phi,W_s$ can switch sign \textit{independently} and the two \textit{together} determine the topological tornado features.
On the other hand, for a scalar field with less information, $P_3$ or the amplitude $|\bP_\mathrm{in}|$, only Eq.~\eqref{eq:spiral_winding} is relevant and suffices to specify the tornado pattern, which will later be cast in the same form as Eq.~\eqref{eq:tornado_winding} from the analytical result.

Table.~\ref{Table:feature_summary} summarizes the correspondence between the three helicity factors and five related aspects in $P_3$ and $\bP_\mathrm{in}$. The dichroic strong/weak response strength of $P_3$ happens with CP light and can be owed to the dipole interband matrix element $\braket{\pm|\hat{\bv}|\mp}$ involving the orbital magnetic moment $\mathcal{M}(\bk)$\cite{Souza2008,Yao2008}. Besides, the $P_3$-tornado displays the extrinsic (intrinsic) helicity factor(s) pinpointedly under CP (LP) light pumping. This is understood as the intrinsic helicities are only transparent under the non-chiral LP light and otherwise overridden by the extrinsic electric field rotation driving the electrons. These features constitute a perfect tomography of the defining helicity parameters of the surface state system and the light-matter interaction, especially given the topological robustness characterized by $W_s$.

However, although tornadoes always exist in the spin-$S_z$ signal $P_3$, their appearance in the vectorial orientation $\phi(\bk)$ of $\bP_\mathrm{in}$ is intriguingly selective.
Considering the nonequilibrium excitations due to the pumping, its winding number two presumably reflects the Berry phase contribution from both particle and hole. Most significantly, other parameters provided, either $W_s$ or $w_\phi$ is nonzero \textit{only} for one type of CP light, making it a novel topological optical activity: dichroic $\mathbb{Z}_2$ topological switching between trivial and nontrivial nonequilibrium responses. Therefore, in addition to the helicity $\Xi_s=\pm1$ dichroic switching of $P_3$, the $\mathbb{Z}_2$ $\bP_\mathrm{in}$-response hints at further possibly interesting ultrafast spintronic applications taking advantage of the two types of all-optical two-state control. 

In fact, the interplay between extrinsic and intrinsic factors %is only in a way hidden in this orientational viewpoint; one 
can also be unmasked through the amplitude $|\bP_\mathrm{in}|$, which exactly follows the response of $P_3$ except a doubled $W_s$, as exemplified in Fig.~\ref{Fig:tornado_illustrate_main}(b).
Unlike the $P_3$-response, aforementioned $\phi$'s radial variation $\mathcal{K}$ is purely locked to $\nu$, giving rise to a stable characterization of the sign of Dirac mass independent of \textit{any} other factors. 
Lastly, in the case of negative spin-orbit coupling that reverses the sign of Fermi velocity $v$, only a sign change of $\bP_\mathrm{in}$ is induced in the in-plane response that does not alter any topological features\cite{Sheng2014,Nie2017}.

The massless side of the phenomena is %, as one can presume on physical grounds, 
presumably simpler: every dichotomous response no longer exists if directly involving the mass sign $\nu$, and only CP light remains active. The purely dichroic tornado in $P_3$ and $|\bP_\mathrm{in}|$ persists. Vanishing mass, however, leads to singular $\pi$-jump in the in-plane $\phi$ along radial direction [e.g., Fig.~\ref{Fig:tornado_main}(a2)]: the tornado trajectory of such domain wall follows the driven dichroic helicity. $\phi$'s variation, i.e., color rotation along the tornado arms, naturally inherits the intrinsic winding sense $\chi$ as in the massive case although the domain wall prevents from completing a quantized winding. The apparent distinction between the massive and massless responses is smoothly connected in the crossover regime $|m|t\sim1$. For instance, tiny amount of magnetic doping ($|m|t\ll1$) follows the massless behavior and the late-time response of finite doping ($|m|t\gg1$) generally obeys the massive response pattern.

\begin{table*}[htb]%The best place to locate the table environment is directly after its first reference in text
\centering
\begin{tabular}{c c | c c | c c}
 %& & \multicolumn{2}{c}{intrinsic}  \\
%\cline{2-5}
%\cmidrule(lr){3-4} %\cmidrule(lr){5-5}
%\cline{3-4} \cline{5-5}
% & & $\chi$ & $\nu$   \\
%\cline{2-4}
%\hline
%\cline{2-5}
%\cmidrule{2-4}  
& & \multicolumn{2}{c|}{massive} & \multicolumn{2}{c}{massless} \\
\cmidrule{3-6} 
%\cline{3-6}
\multirow{3}{*}{\shortstack{normal $P_3$ \& \\ in-plane $|\bP_\mathrm{in}|$}} & \multirow{1}{*}{response strength} & $\quad\chi\nu\tau$ & $\quad\tau=\pm1$ & $\quad$- & $\quad$- \\
%& & \multicolumn{3}{c}{(\qquad\qquad\ \ const. \ \qquad\qquad)}  \\
%\cline{2-4} 
\cmidrule{2-6} 
& \multirow{2}{*}{\shortstack{spiral winding $W_s$\\ ($\times2$ for $|\bP_\mathrm{in}|$)}} & $\quad\chi\nu$ & $\quad\tau=0$  & $\quad$- & $\quad\tau=0$ \\
& & $\quad\tau$ & $\quad\tau=\pm1$  & $\quad\tau$ & $\quad\tau=\pm1$ \\
\cmidrule{1-6} 
 \multirow{5}{*}{\shortstack{in-plane \\ $\phi=\tan^{-1}(\bP_\mathrm{in})$}} & radial $\mathcal{K}=\sgn(\partial_k \phi)$ & $\quad\nu$ & $\quad$-  & $\quad$- & - \\ 
%\cline{2-4} 
\cmidrule{2-6} 
& \multirow{2}{*}{circular winding $w_\phi$} & $\quad0$ & $\quad\chi\nu\tau=0,1$  & \multirow{2}{*}{$\quad\chi^*$} & $\quad\tau=0$ \\
& & $\quad2\chi$ & $\quad\chi\nu\tau=-1$  & \mycomment{$\quad\chi^*$} & $\quad\tau=\pm1$ \\
\cmidrule{2-6} 
& \multirow{2}{*}{spiral winding $W_s=-\mathcal{K} w_\phi$} & $\quad0$ & $\quad\chi\nu\tau=0,1$  & $\quad$- & $\quad\tau=0$ \\
& & $\quad2\tau$ & $\quad\chi\nu\tau=-1$  & $\quad\tau^*$ & $\quad\tau=\pm1$ %\\
\end{tabular}
\caption{\textbf{Correspondence between nonequilibrium topological tornado responses and three system helicity factors} -- intrinsic surface state helicity $\chi=\pm1$, sign of Dirac mass $\nu=\pm1$ or massless case without $\nu$, and extrinsic pump light helicity $\tau=0,\pm1$. Spin-$S_z$ signal $P_3(\bk)$ and in-plane signal amplitude $|\bP_\mathrm{in}|$ show the same dichroism in both the strong or weak ($\pm1$) response strength and the $\bk$-space tornado helicity $\Xi_s=\sgn{W_s}=\pm1$, although spiral winding $W_s$ and hence arm number $\mathcal{R}_s=|W_s|$ are doubled for $|\bP_\mathrm{in}|$. The strength response combines all three factors; $W_s$ is purely driven by extrinsic CP light while it manifests intrinsic factors under LP light. (Massive case $|\bP_\mathrm{in}|$-tornado can be less discernible in Fig.~\ref{Fig:tornado_main} due to obstruction from color, but is otherwise observable when plotted separately. See Fig.~\ref{Fig:tornado_illustrate_main}(b) and \ref{Fig:tornado_main_Pin}.)  In-plane vectorial tornado signal $\bP_\mathrm{in}(\bk)$ contains more information than scalar signals. The azimuthal angle $\phi(\bk)$, encoding the in-plane orientational variation, exhibits a mass-only dependence of $\mathcal{K}$ along the radial direction. Two other related topological winding numbers $w_\phi,W_s$ exist in the massive case and exhibit a $\mathbb{Z}_2$ topological trivial-nontrivial switching for all three factors. $W_s$, common in both scalar and vectorial signals, is driven by CP light ($\tau$) when tornado exists, as it shares the same physical meaning of describing spiral rotation. ($*$) Singular $\pi$-jump domain wall of $\phi$, a double-armed helicity-$\tau$ tornado, disables $\mathcal{K},w_\phi,W_s$ in the massless case; $\chi$ determines $\phi$'s winding sense away from the domain wall for any light polarization.}\label{Table:feature_summary}
\end{table*}

%\mytitle{General linear response}.--
\subsection*{Physical mechanism of tornado}
As seen previously, instead of the possible real transition, virtual excitations giving rise to off-diagonal coherence of electronic density matrix contribute to the tornado formation. 
On top of the ground-state spin-momentum-locked concentric ring-like spin texture, we can intuitively view the optical pumping as producing coherent $\bk$-dependent matrix element and concomitant phase accumulation: the nontrivially correlated phase along the ring rotates the spins to yield the tornado.
This in a way resembles the gas laser, where independent molecules are excited and brought in a correlated nontrivial coherence by the light working as glue. To gain quantitative insight into the nonequilibrium response, we resort to the Keldysh formalism to calculate the crucial $G^<(\bk,t_1,t_2)$ and hence the SARPES signal Eq.~\eqref{eq:I_photo1_main}. In this regard, the linear response is tractable and particularly useful as it captures leading virtual excitations but discards real transitions, given the realistic pumping field is often well within the linear response regime. In addition, since the tornado response is of stable topological nature, the features can persist even beyond as the above relatively larger field calculation confirms. 

The analytical result matches the previous exact calculation in the linear response regime as it should do. For the late-time behavior of our main interest, we can derive an exceptionally simple expression for general two-band systems: $P_0^{(1)}(\varepsilon,\bk,t)\equiv 0$ and
\begin{equation}\label{eq:P_TI_main}
\begin{split}
    \bP^{(1)}(\varepsilon,\bk,t) &=  \frac{2A_0W(k)}{d^2}\left(f_{\varepsilon_{\bk-}} - f_{\varepsilon_{\bk+}}\right) F(\varepsilon)  \tilde{\bP}(\bk,t)
\end{split}
\end{equation}
with $f_{\varepsilon_{\bk\pm}}$ the Fermi function for the upper and lower bands $\varepsilon_{\bk\pm}$.
The vanishing result in the density channel confirms the recovery of stable energy eigenstates after the pump's influence. %, up to possible real transition. 
For the spin channel, the dependence on occupation difference in the two bands indicates the optical inertness of both bands being empty or filled. The energy function in a Gaussian form $F(\varepsilon) = \ee^{-(\varepsilon-d_0(k))^2t_\mathrm{pb}^2}$, where we include $d_0(k)$ for completeness, explains the aforementioned SARPES hot region. The energy range is limited by the probe pulse width; the signal is symmetric with respect to the band midpoint as a result of the interband quantum oscillation.
The momentum envelope function takes a more complex form $W(k)=\sqrt{\frac{\pi}{2}} t_0 \ee^{-2t_0^2(\Omega/2-d(k))^2-t_\mathrm{pb}^2d(k)^2}$ involving both the pump and probe: a disk-like signal centered at $k=0$ can transform to an annulus-like one for large enough $\Omega$ and $t_0$ (see \ref{SM:KeldyshLinear} and \ref{Fig:tornado_main_Wk}). These envelope functions also clarify that the absence or presence of resonant real pumping is inessential to the tornado signal up to minor modification, physically because the interested spin-channel signals rely on the interband coherent dynamics in virtual excitations rather than the real transitions.
Finally, the time-dependent ($\bk$-dependence suppressed and $\partial_j=\partial_{k_j}$)
\begin{equation}\label{eq:P_TI_main1}
\begin{split}
    \tilde{\bP}(\bk,t)&=d\left\{ \left[\tau  (d\,\partial_2\bd-\bd\,\partial_2d) + \bd\times\partial_1\bd\right]\cos{2dt} \right.\\
    &\left.+ \left[- (d\,\partial_1\bd-\bd\,\partial_1d) + \tau \bd\times\partial_2\bd\right]\sin{2dt} \right\}
\end{split}
\end{equation}
\textit{solely} accounts for all the features in Table.~\ref{Table:feature_summary}.
In fact, the scalar $P_3$ or $|\bP_\mathrm{in}|$ admits a generic form
\begin{equation}\label{eq:tornado_form}
    f(\bk)\sin{[2nd(\bk)t+\theta_0-\Theta(\bk)]},
\end{equation}
where $f(\bk)>0$, $n\in\mathbb{Z}_+$, and $\theta_0$ is a constant. While it manifestly originates from the interband coherent oscillation at frequency $2d(\bk)$, the tornado at a given $t$ is made possible since a proper relation between increment of $k$ and $\theta_\bk$ can preserve the argument of sine. Exactly following Eq.~\eqref{eq:tornado_winding}, the spiral winding number $W_s$ is just given by the circular winding $w_\Theta$ of the angle $\Theta(\bk)$.
Representatively, the dichroic $P_3$-tornado reads
\begin{equation}\label{eq:P3_main}
    \tilde{P}_3(\bk,t)=k (d(\bk)+{\chi \tau }m)\sin{[2d(\bk)t+\frac{\pi}{2}-\tau (\theta_\bk+\chi \frac{\pi}{2})]}
\end{equation}
that perfectly explains its appearance in Table.~\ref{Table:feature_summary}. 
The in-plane $\mathbb{Z}_2$ $\phi$-tornado bears a more delicate geometric explanation. %One can decompose at any time $-\tilde{\bP}_\mathrm{in}=\bu+\bv$, where $\bu$ maps the $\bk$-space circle $C_k$ of radius $k$ to an origin-passing doubly traversed circle and $\bv$ translates it. 
The condition in Table.~\ref{Table:feature_summary} exactly specifies whether $\tilde{\bP}_\mathrm{in}$ %, after the translation, 
winds around the origin and hence the trivial or topological winding (see Methods). Correspondingly, $\tilde{\bP}_\mathrm{in}$ crosses the origin only when $m=0$, i.e., the gap closes and hence the singular behavior in massless case, which is the topological transition point along $m$-axis.

To analytically glimpse into possible electronic real transition and nonlinear effects in general, we study as well the special case of a $\delta$-pulse pump, e.g., $\bA(t) = \tilde{A}_0 \delta(t) \hat{x}$, which can account for an LP light ultrashort pump (see Methods).
The nonequilibrium part of SARPES signal reads
\begin{equation}\label{eq:rho_SARPES_pulse_main}
\begin{split}
    \delta P_0(\varepsilon,\bk)&=cE_+(\varepsilon)d(\bk) \\
    \delta\bP(\varepsilon,\bk,t)&=c\left[E_-(\varepsilon)\bd(\bk) + \bar{F}(\varepsilon)\bZ(\alpha,t)\right]
\end{split}
\end{equation} 
where $c=\frac{4\alpha\left(f_{\varepsilon_-} - f_{\varepsilon_+}\right)}{(1+\alpha ^2)^2d^3}$, dimensionless $\alpha=ve\tilde{A}_0/2$ quantifies the deviation from equilibrium, $E_\pm(\varepsilon) = \alpha (d^2-d_y^2) \left[ F_+(\varepsilon) \mp  F_-(\varepsilon)  \right]$, the Gaussian $F_\pm(\varepsilon)=\ee^{-(\varepsilon-\varepsilon_\pm)^2t_\mathrm{pb}^2}$ from the resonant photoemission at two bands, $\bar{F}(\varepsilon)=\ee^{-[(\varepsilon-d_0)^2+d^2]t_\mathrm{pb}^2}$, and $\bZ(\alpha,t)$ in the form of Eq.~\eqref{eq:tornado_form} encodes all linear and nonlinear tornado effects (\ref{SM:delta}). The time-independent $\delta P_0(\varepsilon,\bk)$ describes the result of real pumping from lower $\varepsilon_-$ to higher $\varepsilon_+$. The time-dependent part in the spin channel not only matches Eq.~\eqref{eq:P_TI_main} up to the linear response in $\alpha$, but also suggests the same tornado topology %as linear response 
even deep into the nonlinear regime, which can be confirmed from exact response of short pump pulses. This partially supports the robust observation of tornado topology for moderate strength well beyond linear response regime and 
also hints that general pump pulses can eventually deviate from the linear response prediction of tornado topology at high enough strength.

%phase relaxation $T_2$, energy relaxation $T_1$ (lifetime of excited level): $T_1\gg T_2$

%\mytitle{$\delta$-pulse for LP light}.--
%\section*{$\delta$-pulse for LP light}

%\mytitle{Total magnetization}.--

%\mytitle{Summary}.--
\section*{discussion}

To estimate realistic scales in connection to experiments, we introduce $k_0=\varepsilon_0/v,\mycomment{T_0=2\pi/\varepsilon_0,}\varepsilon_0$ respectively the characteristic scales of wavenumber\mycomment{, time,} and energy. While $\varepsilon_0$ is typically given by the exchange gap $\Delta\sim55\mathrm{meV}$ and hence $k_0\sim 0.03\mathrm{\text{\AA}^{-1}}$ with $v\sim3\times10^5\mathrm{m/s}$ for instance, the driving frequency $\Omega$ can be more important for the gapless or nearly gapless case. 
The dimensionless strength of the pump pulse can be characterized by $\gamma=evA_0/\Omega$, which sensibly relates to the $\delta$-pulse quantity $\alpha=\pi\gamma$. Existing experiments are estimated to fall well within linear response, e.g., $\gamma\sim0.01$\cite{Cacho2015,Jozwiak2016,Reimann2018} (\ref{SM:scale}). 
Exemplifying at $t=0.5\mathrm{ps}$, the tornado arm width $\mycomment{k_\mathrm{arm}}\sim0.01\mathrm{\text{\AA}^{-1}}$.
The femtosecond pump pulse frequency tunes widely from THz to visible; the ultrashort femtosecond probe pulse can provide time duration $0.02\textrm{-}0.5\mathrm{ps}$, energy resolution $5\textrm{-}100\mathrm{meV}$ and momentum resolution $0.004\textrm{-}0.01\mathrm{\text{\AA}^{-1}}$ that are able to observe, given that SARPES signal strength proved to fall well within the experimental reach\cite{Cacho2015,Jozwiak2016,Reimann2018,Lv2019,Sobota2021}. For pump pulse width about the same order of light period $2t_0\sim2\pi/\Omega$ with, e.g., $t_\mathrm{pb}\sim t_0$ and $\Omega\sim\Delta$, an example observation time window after the pump pulse could be $5\textrm{-}150t_0\sim0.2\textrm{-}6\mathrm{ps}$. 
This is feasible in comparison to the experimental estimation of spin relaxation time at the order of $4\textrm{-}15\mathrm{ps}$\cite{Hosur2011,Cacho2015,Iyer2018}. In \ref{SM:relaxation}, taking into account interaction effects, we discuss two relevant and related relaxation time scales: while the energy relaxation time is more easily measurable in experiments, the interband decoherence time plays a more important role in the phenomena of our interest.
Fermi energy inside the gap is not essential since tornado signals persists outside the Fermi ring; finite temperature simply recovers signals inside (\ref{Fig:tornado_main_highermu}). To observe and resolve conspicuous tornado signals in a disk region, shorter $t_0,t_\mathrm{pb}$ and $\Omega$ not very far away from $\Delta$ can help but is not mandatory.

Our results show that the ultrafast spin-resolved response of optically excited topological insulator surface state is an %unexplored but 
exceptionally apt platform %at the interface 
of nonequilibrium topology, coherent quantum dynamics, and light-matter interaction. %Of particular significance are the unique tornado-like response in momentum space, which can dichotomously change winding sense and even switch between topological and trivial depending on various helicity factors from light and material.
%Relevance to 3D Weyl fermion ...
The topology of nonequilibrium spin textures in momentum space will be a new direction in quantum materials. %, which contains rich physics as demonstrated in the present paper. 
Two-dimensional Rashba systems and the generalization to three-dimensional Weyl fermions as well as the spatially nonuniform cases are interesting  problems left for future studies.

%\newpage 
\section*{methods}\label{Sec:Methods}
\subsection*{Model Hamiltonian and time evolution}
We consider a general band electron Hamiltonian
\begin{equation}\label{eq:H_0}
	\hat{H}_0 = \sum_{\bk}\psi^\dagger(\bk) H_0(\bk)\psi(\bk).
\end{equation}
Writing in its tight-binding form for the original lattice model, interaction with a general external electromagnetic field $\bA(\br)$ can be derived from the Peierls substitution %(up to the leading order of $\bA$ and $e>0$)
\begin{equation}
\begin{split}
    & \sum_{\br\br'} \psi^\dag(\br) H_0(\br,\br') \,\ee^{\ii e\mycomment{\frac{e}{\hbar}}\int_{\br'}^{\br}\dd\br'' \cdot\bA(\br'')}  \psi(\br')  \\
    \mycomment{= & \sum_{\br\br'} \psi^\dag(\br) H_0(\br,\br') [1 + \ii e\mycomment{\frac{e}{\hbar}} \br_-\cdot\bA(\br_+)] \psi(\br') \\}
    \approx& \hat{H}_0 + \ii e\mycomment{\frac{e}{\hbar}} \sum_{\bk\bk'} \psi^\dag(\bk) \sum_{\br\br'} \ee^{\ii (\bk_-\cdot\br_+ + \bk_+\cdot\br_-)} H_0(\br_-)\times\\ 
    &\br_-\cdot\bA(\br_+) \psi(\bk') \\
    \mycomment{=& \hat{H}_0 + e\mycomment{\frac{e}{\hbar}} \sum_{\bk\bk'} \psi^\dag(\bk) \partial_{\bk_+} \cdot \left(\sum_{\br_-\br_+} \ee^{\ii (\bk_-\cdot\br_+ + \bk_+\cdot\br_-)} \bA(\br_+) H_0(\br_-) \right) \psi(\bk') \\}
    =& \hat{H}_0 + e\mycomment{\frac{e}{\hbar}} \sum_{\bk\bk'} \psi^\dag(\bk)  \partial_{\bk_+} H_0(\bk_+) \cdot \bA(\bk_-) \psi(\bk') \\
\end{split}
\end{equation}
where we denote $\br_-=\br-\br',\br_+=\frac{\br+\br'}{2}$ and similarly for $\bk_\pm$. We use the fact that $H_0(\br,\br')$ is periodic and approximate the Peierls phase by the midpoint valued $\bA$ accumulated along the \mycomment{shortest }path connecting the two sites, which is justified as the long-wavelength electromagnetic field is slowly varying at atomic scales. 
Therefore, in the optical long-wavelength limit of a spatially uniform time-dependent $\bA(t)$, we obtain Eq.~\eqref{eq:H_main}.

%\subsection*{Time evolution}
The unitary time evolution can be performed via the equation of motion (EOM) of the column field vector $\psi(t)$ in the Heisenberg picture
\begin{equation}\label{eq:HEOM}
    \ii\mycomment{\hbar\,}\dot\psi(t)=[\psi(t), \hat{H}^\mathrm{H}(t)],
\end{equation}
where $\hat{H}^\mathrm{H}(t)=H_{\alpha\beta}(t)\psi_\alpha^\dag(t)\psi_\beta(t)$ and we neglect $\bk$-dependence for brevity. As required by the unitary time evolution of any operator $\psi_\alpha(t)=\hat{U}(t)\psi_\alpha\hat{U}^\dag(t)$, the equal-time canonical commutation relation should always hold
\begin{equation}\label{eq:CCR}
\begin{aligned}
&\{\psi_\alpha(t),\psi_\beta^\dag(t)\} = \delta_{\alpha\beta}\\ &\{\psi_\alpha^\dag(t),\psi_\beta^\dag(t)\} = \{\psi_\alpha(t),\psi_\beta(t)\} = 0.
\end{aligned}
\end{equation}
% We adopt the \textit{ansatz} that attributes operator time dependence to a coefficient matrix
% \begin{equation}\label{eq:iEOMansatz}
%     \psi_\alpha(t) = B_{\alpha\beta}(t) \psi_\beta,
% \end{equation}
% which leads to a closed solution form for a quadratic Hamiltonian. In the present choice of the dynamical operators, we have the natural initial condition 
% \begin{equation}\label{eq:iEOM_IC}
%     B_{\alpha\beta}(-\infty)=\delta_{\alpha\beta}.
% \end{equation}
% From Eq.~\eqref{eq:HEOM}, we can derive an apparently nonlinear matrix EOM
% \begin{equation}\label{eq:B_EOM}
% \begin{split}
%     \ii\mycomment{\hbar}\dot B(t) = B(t)M(t)
% \end{split}
% \end{equation}
% where $M(t)=B^\dag(t)H(t)B(t)$ is Hermitian and we use %the equilibrium version of Eq.~\eqref{eq:CCR}, i.e., 
% the canonical commutation relation for the time-independent Schr\"odinger operators.
% To ensure the validity of the ansatz, one can now verify the unitarity and hence the general Eq.~\eqref{eq:CCR} by the invariant $B(t)B^\dag(t) = \mathds{I}$ as a consequence of the evolution, which can be proved from Eqs.~\eqref{eq:iEOM_IC}\eqref{eq:B_EOM}. Under this situation we reduce Eq.~\eqref{eq:B_EOM} to the matrix EOM
We adopt the \textit{ansatz} that attributes operator time dependence to a coefficient matrix $\psi_\alpha(t) = B_{\alpha\beta}(t) \psi_\beta$,
which leads to a closed solution form for a quadratic Hamiltonian. In the present choice of the dynamical operators, we have the natural initial condition $B_{\alpha\beta}(-\infty)=\delta_{\alpha\beta}$. From \eqref{eq:HEOM}, we can derive an apparently nonlinear matrix EOM
\begin{equation}\label{eq:B_EOM}
\begin{split}
    \ii\mycomment{\hbar}\dot B(t) = B(t)M(t)
\end{split}
\end{equation}
where $M(t)=B^\dag(t)H(t)B(t)$ is Hermitian and we use %the equilibrium version of \eqref{eq:CCR}, i.e., 
the canonical commutation relation for the time-independent Schr\"odinger operators.
To ensure the validity of the ansatz, one can now verify the unitarity and hence the general \eqref{eq:CCR} by the invariant $B(t)B^\dag(t) = \mathds{I}$ as a consequence of the evolution, which can be proved from the initial condition and \eqref{eq:B_EOM}. Under this situation we reduce \eqref{eq:B_EOM} to the matrix EOM
\begin{equation}\label{eq:B_EOM1}
\begin{split}
    \ii\mycomment{\hbar}\dot B(t) = H(t)B(t)
\end{split}
\end{equation}
that fully determines the time-dependent system and can be solved numerically. 

The double-time %removal 
Green's function with nonequilibrium information, introduced in the main text, can be related to 
\begin{equation}
    G^<(\bk,t_1,t_2) = B(\bk,t_1) G_0^<(\bk) B^\dag(\bk,t_2) 
\end{equation}
with the equilibrium Green's function
\begin{equation}\label{eq:G^<0mat}
\begin{split}
    G_0^<(\bk) &= \ii \sum_{a=\pm} f_{\varepsilon_{\bk a}} |\bk a\rangle \langle \bk a| \\
    &= \ii \frac{(\ee^{- d_0\beta}+\cosh{d\beta})\sigma_0 - \sinh{d\beta}\,\hat{\bd}\cdot\bsigma}{2\cosh{d_0\beta}+2\cosh{d\beta}}
\end{split}
\end{equation}
specified from the band basis $|\bk a\rangle$ using the Fermi distribution $f_{\bk a}=(\ee^{\beta(\varepsilon_{\bk a}-\mu)}+1)^{-1}$ and given in Pauli decomposition form.

\subsection*{Keldysh response theory}
In the time-contour (forward '$+$' branch and backward '$-$' branch) formalism of nonequilibrium Green's function, we have the Green's function matrix
\begin{equation}
\hat{G}=\begin{bmatrix}
G^{++} & G^{+-} \\
G^{-+} & G^{--} 
\end{bmatrix}
=\begin{bmatrix}
G^{\mathds{T}} & G^< \\
G^> & G^{\tilde{\mathds{T}}} 
\end{bmatrix}
\end{equation}
and the Keldysh rotated one
\begin{equation}\label{eq:KeldyshRotation}
    \check{G}=R\hat{G}R^\dag=\begin{bmatrix}
0 & G^\mathrm{a} \\
G^\mathrm{r} & G^\mathrm{k}
\end{bmatrix}
\end{equation}
with $R=\frac{1}{\sqrt{2}}\begin{bmatrix}
1 & -1 \\
1 & 1
\end{bmatrix}$. The Dyson equation $G = G_0(1+ \Sigma G)$ holds for both cases where Keldysh-space matrix multiplication and argument convolution is understood. The corresponding self-energy matrices in the Keldysh space read in the present case
\begin{equation}\label{eq:Sigma0}
    \hat{\Sigma}(\bk,t;\bk',t') = \Sigma_0\,\sigma_3,\quad \check{\Sigma}(\bk,t;\bk',t') = \Sigma_0\,\sigma_1
\end{equation}
with $\Sigma_0 = H'(\bk,t)\delta(\bk-\bk')\delta(t-t')$ and $H'(\bk,t)$ the pumping interaction Hamiltonian we derived.
From the exact Dyson equation of $G^<$ 
\begin{equation}\label{eq:G^<identity}
    G^< = (1+G^\mathrm{r}\Sigma^\mathrm{r}) G_0^< (1+\Sigma^\mathrm{a}G^\mathrm{a}) + G^\mathrm{r} \Sigma^< G^\mathrm{a},
\end{equation}
we can obtain the linear response
\begin{equation}\label{eq:G^<_1}
    G_1^{<} = G_0^{<} \Sigma_0G_0^\mathrm{a} + G_0^\mathrm{r} \Sigma_0G_0^{<}.
\end{equation}

As per our purpose, we evaluate $\mathscr{G}_i = \mathrm{Tr}[G^<_1\sigma_i]$ and derive the analytical form
\begin{equation}\label{eq:KeldyshGra<}
\begin{split}
    &\mathscr{G}_i(\bk,t_1,t_2) \\
    =& \int_{-\infty}^{t_2} \dd t A_\kappa( t) Y_i^\kappa(\bk,t_+,t_-) - \int_{-\infty}^{t_1} \dd t A_\kappa( t) Z_i^\kappa(\bk,t_+,t_-),
\end{split}
\end{equation}
where $\kappa=1,2$, $t_+ = t_1+t_2-2t,t_- = t_1-t_2$, and
\begin{equation}\label{eq:XYZmualpha_1storder}
\begin{split}
    & Y_i^\kappa(\bk,t_+,t_-) 
    = -\frac{\ee^{-(d_0-\mu)\beta} \, X_i^\kappa + X_i^\kappa\big|_{ t_\pm\rightarrow t_\pm - \ii\beta} }{\cosh{(d_0-\mu)\beta}+\cosh{d\beta}} \\
    & Z_i^\kappa(\bk,t_+,t_-) 
    = -\frac{\ee^{-(d_0-\mu)\beta} \, X_i^\kappa + X_i^\kappa\big|_{ t_\pm\rightarrow t_\pm \pm \ii\beta} }{\cosh{(d_0-\mu)\beta}+\cosh{d\beta}} 
\end{split}
\end{equation}
with $d\ee^{\ii t_-d_0}X_i^\kappa(\bk,t_+,t_-)$ given by $d ( \partial^\kappa d_0 \cos{d t_-} - \ii \hat{\bd}\cdot\partial^\kappa\bd \sin{d t_-} )$ when $i=0$
and 
$-\ii d_i\partial^\kappa d_0 \sin{d t_-} + \left( \bd \times \partial^\kappa \bd  \right)_i \sin{d t_+}  + d_i \hat{\bd}\cdot\partial^\kappa\bd \cos{d t_-} + (  d \partial^\kappa d_i - d_i \hat{\bd} \cdot \partial^\kappa \bd  ) \cos{d t_+} $ when $i=1,2,3$. Now Eq.~\eqref{eq:KeldyshGra<} can be evaluated analytically using a simple special function
\begin{equation}\label{eq:Ierf1}
\begin{split}
        &I(\omega,a,T) \\
        =& \frac{1}{2} \int_{-\infty}^T \dd\tau \: \ee^{-\frac{\tau^2}{2t_0^2}} \ee^{\ii[\omega\tau + a(t-\tau)]} \\
        =& \sqrt{\frac{\pi}{8}}t_0\, \ee^{- \frac{t_0^2}{2}(\omega- a)^2}  \ee^{\ii at} \, \left(1+\mathrm{Erf}(\frac{T-\ii(\omega- a)t_0^2}{\sqrt{2}t_0})\right)
\end{split}
\end{equation}
with $\omega=\pm\Omega,a=2d,T=t_{1,2}$. We present the detailed relation in \ref{SM:KeldyshLinear}. This fully analytical theory of the double-time removal Green's function matches the exact numerical time evolution better and better towards the linear response regime, e.g., when $A_0<0.05$.

To elucidate the tornado responses, we especially focus on the late-time behavior where the error function in Eq.~\eqref{eq:Ierf1} approaches unity when $T\gg t_0$.
Now Eq.~\eqref{eq:I_photo1_main} can be further evaluated analytically. We arrive at the most general form of the late-time SARPES signal for a two-band model
\begin{equation}\label{eq:P_general}
\begin{split}
    &P_0^{(1)}(\varepsilon,\bk,t)\equiv 0 \\
    &\bP^{(1)}(\varepsilon,\bk,t) = \frac{2A_0}{d}\left(f_{\varepsilon_-} - f_{\varepsilon_+}\right) F(\varepsilon) \times \\
    &\left\{ \left[\tau W_s (d\,\partial_2\bd-\bd\,\partial_2d) + W_c\,\bd\times\partial_1\bd\right]\cos{2dt} \right.\\
    &\left.+ \left[-W_c (d\,\partial_1\bd-\bd\,\partial_1d) + \tau W_s\,\bd\times\partial_2\bd\right]\sin{2dt} \right\}
\end{split}
\end{equation}
with $W_{c,s} =\sqrt{\frac{\pi}{2}} t_0 \ee^{-d^2t_\mathrm{pb}^2} \sum_{a=\pm}a^x\,\ee^{-\frac{t_0^2}{2}(a\Omega-2d)^2}$ where $x=0,1$ respectively for $W_{c,s}$. Without affecting any topological features, one can approximate $W=W_{c,s}=\sqrt{\frac{\pi}{2}} t_0 \ee^{-\frac{t_0^2}{2}(\Omega-2d)^2-d^2t_\mathrm{pb}^2}$ and reach Eq.~\eqref{eq:P_TI_main}.

\subsection*{Topological tornado response}
The topological tornado information in Eq.~\eqref{eq:P_TI_main1} can be seen through simplification towards the general form Eq.~\eqref{eq:tornado_form} for the specific scenarios, in a similar manner as Eq.~\eqref{eq:P3_main}.
For instance, when $\tau =0$, we instead have ($v=1$)
\begin{equation}\label{eq:z_winding_main}
\begin{split}
    &\tilde{P}_3(\bk,t)=
    \sqrt{m^2k_x^2+d^2k_y^2}\\ &\times\sin{[2dt+\frac{\pi}{2}-\nu (\chi \arctan(|m|k_x,dk_y)+\frac{\pi}{2})]}.
\end{split}
\end{equation}
Other situations are discussed in \ref{SM:topo}. 

Now we briefly sketch the proof of the $\mathbb{Z}_2$ orientational $\bP_\mathrm{in}$-tornado. We decompose 
$-\tilde{\bP}_\mathrm{in}=\bu + \bv$
where 
\begin{equation}\label{eq:u&v_split_main}
    \bu = \left(\bk_{\tau }\cdot\hat{\bq}\right)\bk_{\chi },\quad \bv = m\begin{pmatrix} d+{\chi \tau } m & \\ & {\chi \tau } d+m \end{pmatrix}
    \hat{\bq}
\end{equation}
with $\bk_\pm=(\pm k_x,k_y),\hat{\bq} = (\cos{2dt},\sin{2dt})$. Given $k$, i.e., a circle $C_k$ on the 2D $\bk$-plane, $\bv$ is a constant vector field. While $\bu$ is oriented parallel to the radial direction of $\hat{\bk}_{\chi }$ it vanishes at two diametrically opposite points on $C_k$ where $\bk_{\tau }\perp\hat{\bq}$. In fact, the vector field $\bu$ maps $C_k$ to a new trajectory, a circle $\mathcal{C}_k$ that is doubly and $\chi $-clockwisely traversed and also passes the origin twice. For the translated circular trajectory $\mathscr{C}_k$ of $\tilde{\bP}_\mathrm{in}$, a key observation is that as long as $m\neq0,k>0$
\begin{equation}
    \begin{cases}
    \textrm{$\tilde{\bP}_\mathrm{in}=\bzero$ lies outside $\mathscr{C}_k$} & \textrm{${\tau }=0$ or ${\chi \tau }\nu =1$} \\
    \textrm{$\tilde{\bP}_\mathrm{in}=\bzero$ lies inside $\mathscr{C}_k$} & {\chi \tau }\nu =-1
    \end{cases},
\end{equation}
which immediately dictates the $\mathbb{Z}_2$ response.

To see the robust correspondence to the sign of mass $\sgn{(\partial_k\phi)}=\nu $ in the in-plane orientational signal $\phi(\bk)$, we rely on the one-form $\dd\phi = \frac{1}{|\tilde{P}_\mathrm{in}|^2}(\tilde{P}_x \dd \tilde{P}_y -\tilde{P}_y \dd \tilde{P}_x)$. In \ref{SM:topo}, we prove that $\frac{2d}{km}(\tilde{P}_x \partial_k \tilde{P}_y -\tilde{P}_y \partial_k \tilde{P}_x)>0$ when $t>\frac{1}{2|m|}$ in general holds.

\subsection*{\texorpdfstring{$\delta$}{delta}-pulse for LP light}
Note that $\delta$-pulse is not feasible to describe a CP light pulse since $\delta(t)$ automatically picks out one particular Hamiltonian at $t=0$. For the LP light polarized along $\hat{x}$, we consider the Hermitian evolution generator $S=B^\dag(0^-)H(0)B(0^-)$ for Eq.~\eqref{eq:B_EOM} for an infinitesimal pulse duration $\Delta t$, leading to 
\begin{equation}
\begin{split}
    S\frac{\Delta t}{2} |_{\Delta\rightarrow0,\delta(t)\Delta t\rightarrow1}
     =\frac{\alpha}{v} B^\dag(0^-)\partial^1H_0B(0^-).
\end{split}
\end{equation}
It is crucial to make the $\delta$-pulse evolution \textit{unitary}, which can be achieved via the Pad\'{e} approximant that divides the pulse into two parts, i.e., $t<0$ and $t>0$ parts. For the $\delta$-pulse, it suffices to apply the $R_{1,1}$ approximant\cite{NumericalRecipes}
\begin{equation}
\begin{split}
    B(0^+) &= B(0^-) (I-\ii S\frac{\Delta t}{2})(I+\ii S\frac{\Delta t}{2})^{-1} .
\end{split}
\end{equation}
After the pulse, we have the time evolution $B(t) = U(t) B(0^+)$
with 
\begin{equation}
\begin{split}
    U(t)=\ee^{- \ii H_0 t}= \ee^{-\ii d_0 t} \left( \cos{d t} \, \sigma_0- \ii\sin{d t} \,\hat{\bd}\cdot\bsigma \right)
\end{split}
\end{equation}
since the time-dependent drive is off. Then one can derive \eqref{eq:rho_SARPES_pulse_main}. See \ref{SM:delta}.

%\subsection*{Scale estimation}

\section*{acknowledgments}
X.-X.Z. appreciates helpful discussion with L. Schwarz, Y. Fan, I. Belopolski, A. F. Kemper and J. K. Freericks. This work was supported by JSPS KAKENHI (No.~18H03676) and JST CREST (Nos.~JPMJCR16F1 \& JPMJCR1874). X.-X.Z. was partially supported by the Riken Special Postdoctoral Researcher (SPDR) Program. 
\mycomment{\Yinyang}

% \section*{author contributions}
% X.X.-Z. carried out the model calculation and analyzed the data together with N.N.. N.N. supervised the project. Both authors discussed the results and wrote the manuscript.

%\nocite{*}

\bibliography{reference.bib}
  % The references (bibliography) information are stored in the file named "Bibliography.bib"

\let\addcontentsline\oldaddcontentsline% Restore \addcontentsline

% \clearpage 
% \onecolumngrid
% \newpage
% %%%%%%%%%% Merge with supplemental materials %%%%%%%%%%
% %%%%%%%%% Prefix a "S" to all equations, figures, tables and reset the counter %%%%%%%%%%
% %\appendix
% \setcounter{equation}{0}
% \setcounter{figure}{0}
% \setcounter{table}{0}
% \setcounter{page}{1}
% %\makeatletter
% \renewcommand{\theequation}{S\arabic{equation}}
% \renewcommand{\thefigure}{S\arabic{figure}}
% \renewcommand{\thetable}{S\arabic{table}}
% \renewcommand{\theHtable}{Supplement.\thetable}
% \renewcommand{\theHfigure}{Supplement.\thefigure}
% \renewcommand{\thesection}{Supplementary Note~\arabic{section}}
% \renewcommand{\bibnumfmt}[1]{[S#1]}
% \renewcommand{\citenumfont}[1]{S#1}
% %%%%%%%%% Prefix a "S" to all equations, figures, tables and reset the counter %%%%%%%%%%

%\listoffigures

\clearpage
\onecolumngrid
\newpage
{
	\center \bf \large 
	Supplementary Information \\
	%\large for ``Non-Hermitian exceptional Landau quantization in electric circuits"\vspace*{0.1cm}\\ 
	\large for ``\newtitle"\vspace*{0.1cm}\\ 
	\vspace*{0.5cm}
	%\newauthor
}
% \begin{center}
%     %\getauthor \\
% 	Xiao-Xiao Zhang and Marcel Franz\\
% 	\vspace*{0.15cm}
% 	\small{\textit{Department of Physics and Astronomy \& Stewart Blusson Quantum Matter Institute, University of British Columbia, Vancouver, BC, V6T 1Z4 Canada}}\\
% 	\vspace*{0.25cm}	
% \end{center}

%\twocolumngrid	

\tableofcontents

% %\clearpage
% %\appendix
% \setcounter{equation}{0}
% \setcounter{figure}{0}
% \setcounter{table}{0}
% \setcounter{page}{1}
% %\renewcommand{\theequation}{S\arabic{equation}}
% \renewcommand{\thefigure}{S\arabic{figure}}
% \renewcommand{\bibnumfmt}[1]{[S#1]}
% %\renewcommand{\citenumfont}[1]{S#1}

%%%%%%%%%% Merge with supplemental materials %%%%%%%%%%
%%%%%%%%% Prefix a "S" to all equations, figures, tables and reset the counter %%%%%%%%%%
%\appendix
\setcounter{equation}{0}
\setcounter{figure}{0}
\setcounter{table}{0}
\setcounter{page}{1}
%\makeatletter
\renewcommand{\theequation}{S\arabic{equation}}
\renewcommand{\thefigure}{S\arabic{figure}}
\renewcommand{\thetable}{S\arabic{table}}
\renewcommand{\theHtable}{Supplement.\thetable}
\renewcommand{\theHfigure}{Supplement.\thefigure}
\renewcommand{\thesection}{Supplementary Note \arabic{section}}
\renewcommand{\bibnumfmt}[1]{[S#1]}
\renewcommand{\citenumfont}[1]{S#1}
%%%%%%%%% Prefix a "S" to all equations, figures, tables and reset the counter %%%%%%%%%%
\newpage
\section*{Supplementary Data Figures}\label{SM:Figs}
\begin{figure*}[hbt]
\includegraphics[width=17.9cm]{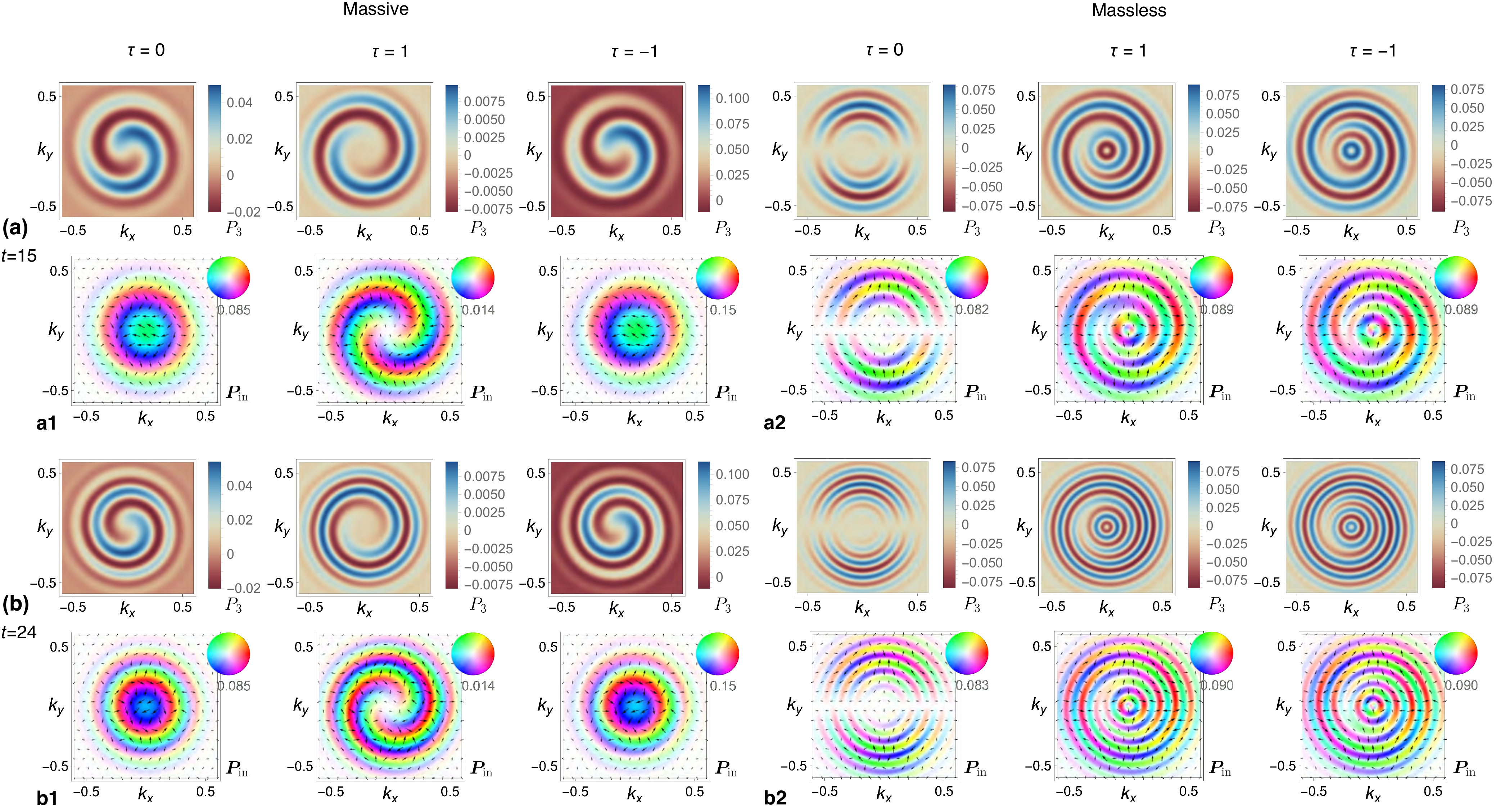}
\caption{\textbf{Nonequilibrium tornado-like responses when $\chi=-1$.} Same SARPES signals (equilibrium response subtracted) as Fig.~\ref{Fig:tornado_main} with reversed surface state helicity $\chi=-1$. Other parameters same as Fig.~\ref{Fig:tornado_main}.}\label{Fig:tornado_main_rvs_chi}
\end{figure*}

\begin{figure*}[hbt]
\includegraphics[width=9cm]{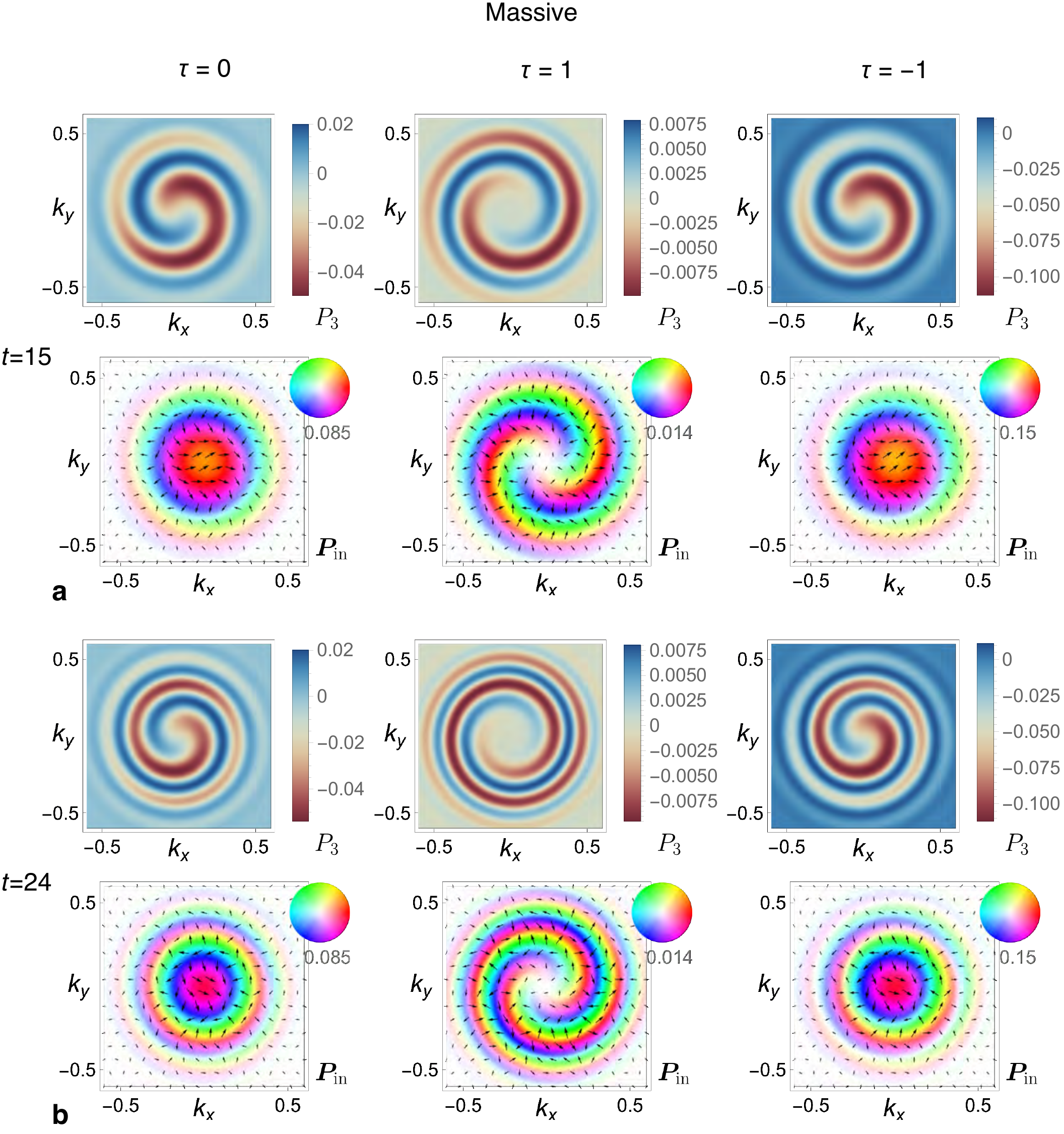}
\caption{\textbf{Nonequilibrium tornado-like responses when $\nu=-1$.} Same SARPES signals (equilibrium response subtracted) as Fig.~\ref{Fig:tornado_main} with reversed surface state sign of mass $\nu=-1$. Other parameters same as the massive case in Fig.~\ref{Fig:tornado_main}.}\label{Fig:tornado_main_rvs_nu}
\end{figure*}

\begin{figure*}[hbt]
\includegraphics[width=9cm]{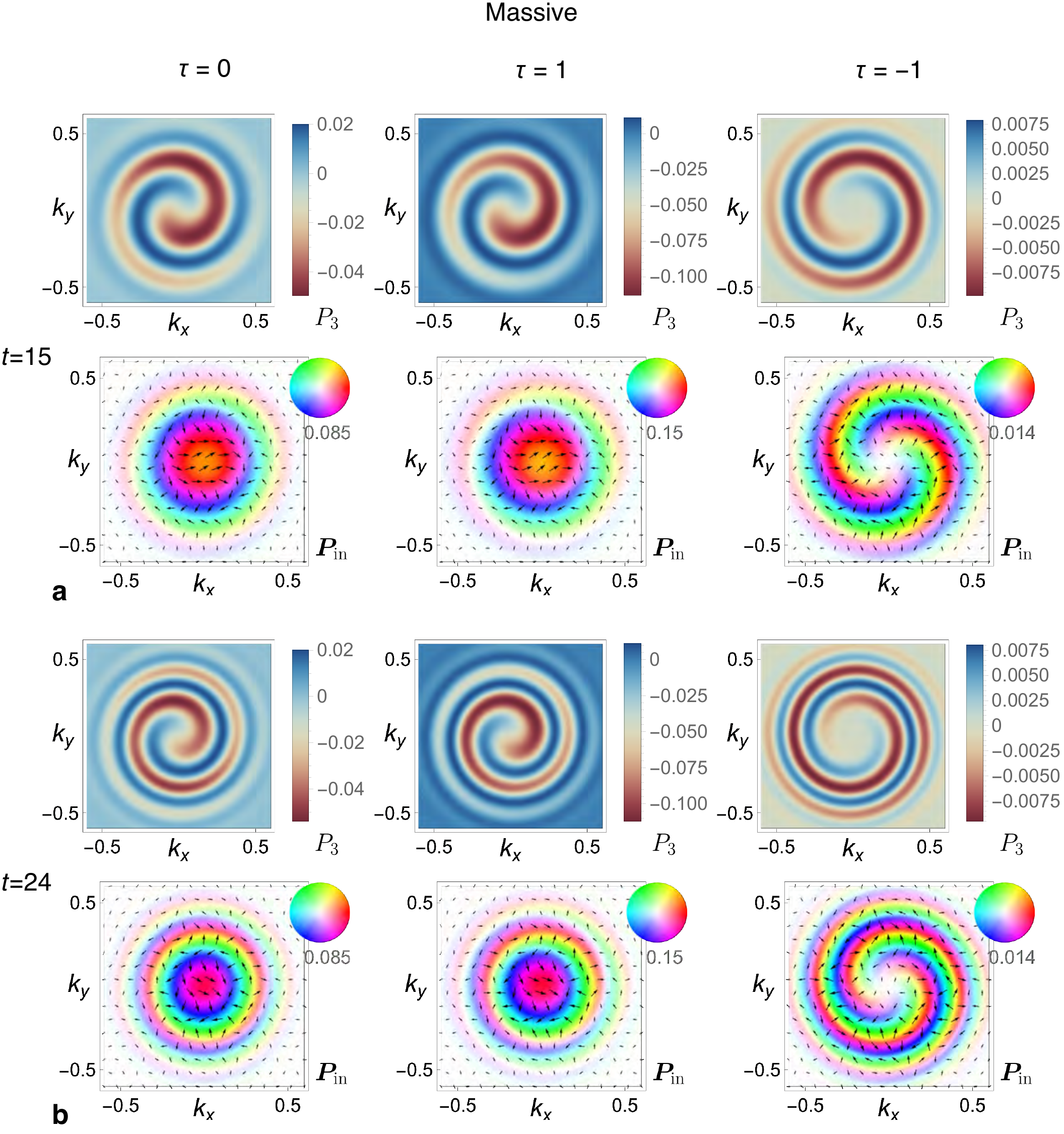}
\caption{\textbf{Nonequilibrium tornado-like responses when $\chi=-1$ and $\nu=-1$.} Same SARPES signals (equilibrium response subtracted) as Fig.~\ref{Fig:tornado_main} with both reversed surface state helicity $\chi=-1$ and reversed surface state sign of mass $\nu=-1$. Other parameters same as the massive case in Fig.~\ref{Fig:tornado_main}.}\label{Fig:tornado_main_rvs_chi_nu}
\end{figure*}

\begin{figure*}[hbt]
\includegraphics[width=17.9cm]{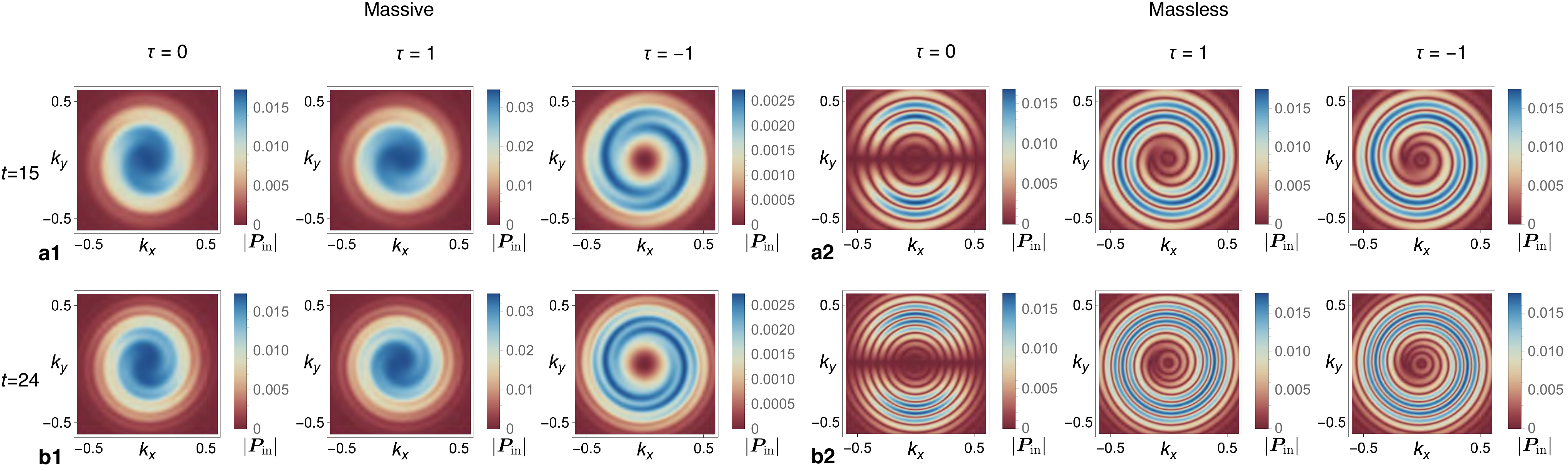}
\caption{\textbf{Nonequilibrium tornado-like responses in the in-plane amplitude.} Similar SARPES signals (equilibrium response subtracted) as Fig.~\ref{Fig:tornado_main} for the amplitude $|\bP_\mathrm{in}|$. Calculation at $A_0=0.02$ and other parameters same as Fig.~\ref{Fig:tornado_main}. Essential tornado response features follow the spin-$S_z$ signal $P_3$ except that the spiral winding number $W_s$ and the arm number $\mathcal{R}_s$ are doubled.}\label{Fig:tornado_main_Pin}
\end{figure*}

\begin{figure*}[hbt]
\includegraphics[width=17.9cm]{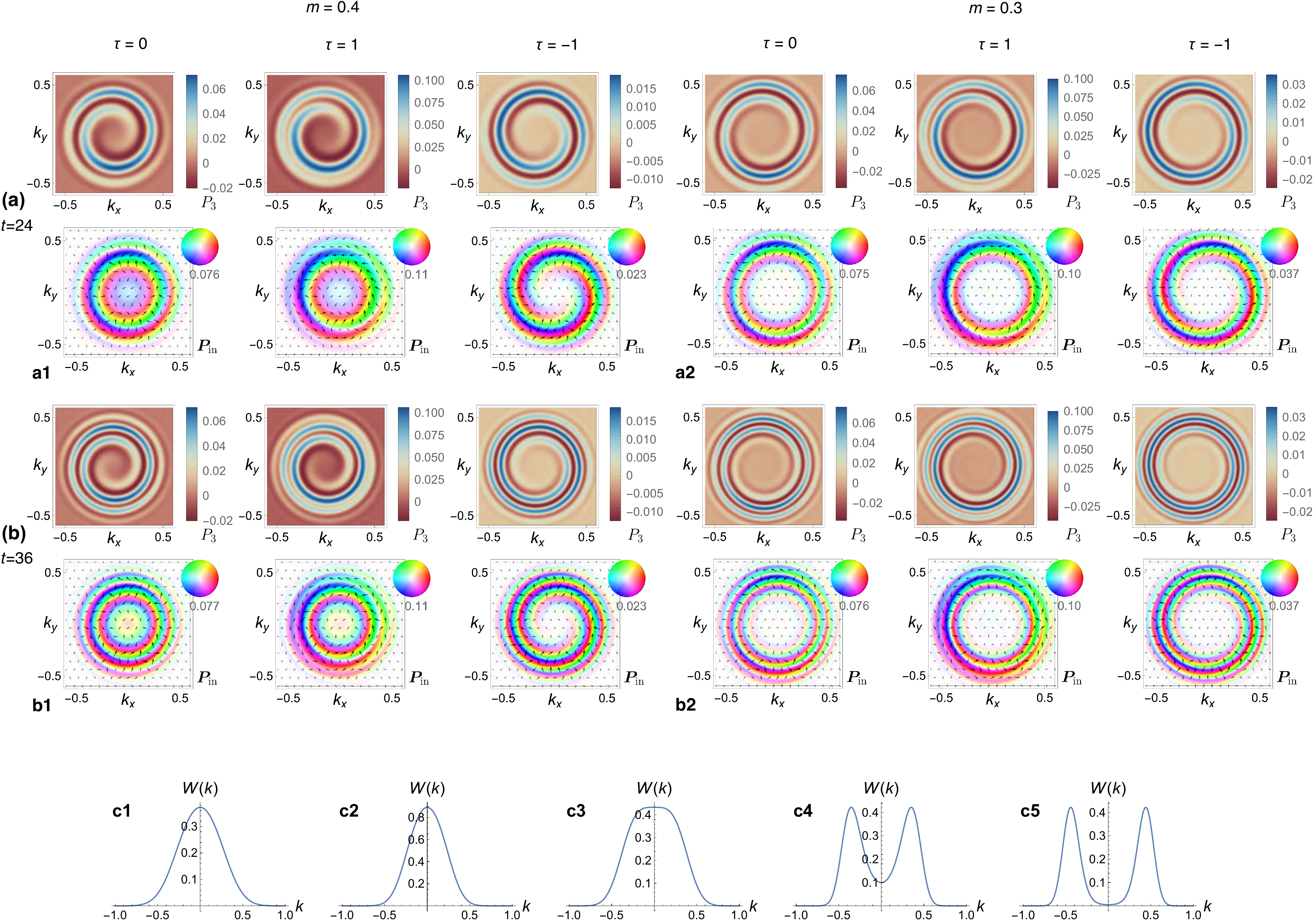}
\caption{\textbf{Momentum envelope shape of nonequilibrium tornado-like responses.} Same SARPES signals (equilibrium response subtracted) as Fig.~\ref{Fig:tornado_main} for a longer pump pulse $t_0=2t_\mathrm{pb}=6$ with different masses $m=0.4$ (a1,b1) and $m=0.3$ (a2,b2). Time snapshots at (a) $t=24$ (b) $t=36$. Other calculation parameters same as Fig.~\ref{Fig:tornado_main}. Compared with the disk-like momentum envelope in Fig.~\ref{Fig:tornado_main}, here (a) and (b) show more annulus-like momentum envelope distribution.
(c) Profile of the analytical momentum envelope function $W(k)$ for five example parameter sets. Parameters not mentioned are the same as (c3). (c1) Shorter pump pulse $t_0=1.5$; (c2) smaller pumping frequency at marginal resonance $\Omega=\Delta=0.8$; (c3) massive case of Fig.~\ref{Fig:tornado_main}; (c4) longer pump pulse $t_0=6$ for (a1,b1); (c5) longer pump pulse $t_0=6$ and smaller mass $m=0.3$ for (a2,b2). Analytical result in (c) well captures the exact simulations.
}\label{Fig:tornado_main_Wk}
\end{figure*}

\begin{figure*}[hbt]
\includegraphics[width=17.9cm]{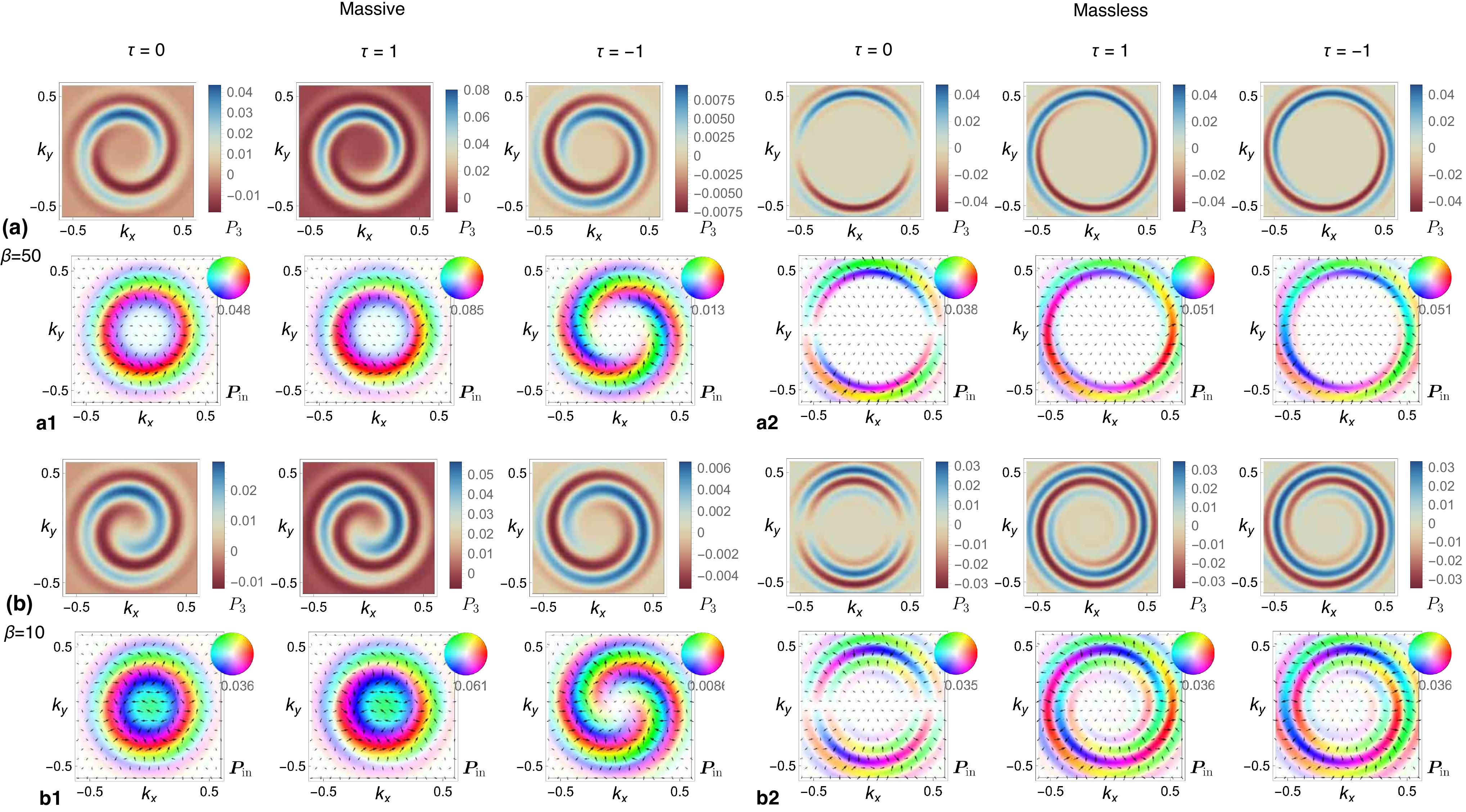}
\caption{\textbf{Fermi energy and temperature dependence.} Same SARPES signals (equilibrium response subtracted) as Fig.~\ref{Fig:tornado_main} for a higher Fermi energy $\mu=0.48$ crossing the upper band at $t=15$. Two different temperatures (a) $\beta=50$ and (b) $\beta=10$ are considered. Other parameters same as Fig.~\ref{Fig:tornado_main}. (a) The inactive region inside the Fermi ring can be clearly seen, which is smaller in the gapful case. Outside the Fermi ring, tornado features remain intact. (b) Higher temperature can render the region inside the Fermi ring active in the optical nonequilibrium process, but does not affect the essential tornado features.}\label{Fig:tornado_main_highermu}
\end{figure*}
\clearpage

\section{SARPES formalism}\label{SM:SARPES}
The time-resolved SARPES intensity in the main text can be derived by generalizing Ref.~\cite{Freericks2009_SI} and we mainly follow the notation therein. %The photocurrent operator reads $\bJ(\bR)\simeq\frac{\bk}{m}c^\dag_{\bk,\bR}c_{\bk,\bR}$ with $c^\dag_{\bk,\bR}=\sum_{\nu'\sigma'\bk'}\phi_{\bk,\bR}(\nu',\sigma',\bk')c^\dag_{\nu'\sigma'\bk'}$.
The part of probe pulse interaction at time $t'$ corresponding to absorbing a photon of momentum $\bq$ and frequency $\omega_\bq$ is 
\begin{equation}
    \mathcal{H}_\mathrm{pb}(t')=\sum_{\nu\nu'\sigma\bk}s(t')\,\ee^{-\ii\omega_\bq t'}M_\bq(\nu,\nu',\sigma,\bk,t')c^\dag_{\nu'\sigma\bk+\bq}c_{\nu\sigma\bk}a_\bq
\end{equation}
where probe pulse profile $s(t')$ is given in the main text, $\sigma$ denotes spin that is preserved, $\nu$ refers to any other quantum number, $c_{\nu\sigma\bk},a_\bq$ are respectively the electron and photon annihilation operator, and $M_\bq$ is the interaction matrix element. Evaluating the photocurrent expectation value, one can extract the SARPES intensity detected from the probe pulse centered around $t$ that is encoded in $s(t')$
\begin{equation}
\begin{split}
    P(\nu\bk\nu'\bk'\sigma_1\sigma_2)
    =& \sum_{\nu_1\nu_1'\bk_1\nu_2\nu_2'\bk_2\sigma\sigma'} \int\dd t_1 \dd t_2 M^*_\bq(\nu_2,\nu_2',\sigma_2,\bk_2,t_2) M_\bq(\nu_1,\nu_1',\sigma_1,\bk_1,t_1)s(t_1)s(t_2)\ee^{\ii\omega_\bq(t_2-t_1)} W
\end{split}
\end{equation}
where $W = \langle c^\dag_{\nu_2\sigma_2\bk_2}(t_2)c_{\nu_2'\sigma_2\bk_2+\bq}(t_2) c^\dag_{\nu'\sigma'\bk'}(t')  c_{\nu\sigma\bk}(t') 
    c^\dag_{\nu_1'\sigma_1\bk_1+\bq}(t_1)c_{\nu_1\sigma_1\bk_1}(t_1) \rangle \mycomment{\\
    & = \left\langle c^\dag_{\nu_2\sigma_2\bk_2}(t_2)c_{\nu_1\sigma_1\bk_1}(t_1)\right\rangle_{\mathcal{H}} \delta_{\nu_2'\nu'}\delta_{\sigma_2\sigma'}\delta_{\bk_2+\bq,\bk'} \delta_{\nu\nu_1'}\delta_{\sigma\sigma_1}\delta_{\bk,\bk_1+\bq} \ee^{\ii[(\varepsilon_{\nu'\bk'}-\mu)(t-t_2)-(\varepsilon_{\nu\bk}-\mu)(t-t_1)]}}$
% \begin{equation}
% \begin{split}
%     W &= \left\langle c^\dag_{\nu_2\sigma_2\bk_2}(t_2)c_{\nu_2'\sigma_2\bk_2+\bq}(t_2) c^\dag_{\nu'\sigma'\bk'}(t') \right.\\
%     & \left. c_{\nu\sigma\bk}(t') c^\dag_{\nu_1'\sigma_1\bk_1+\bq}(t_1)c_{\nu_1\sigma_1\bk_1}(t_1) \right\rangle \mycomment{\\
%     & = \left\langle c^\dag_{\nu_2\sigma_2\bk_2}(t_2)c_{\nu_1\sigma_1\bk_1}(t_1)\right\rangle_{\mathcal{H}} \delta_{\nu_2'\nu'}\delta_{\sigma_2\sigma'}\delta_{\bk_2+\bq,\bk'} \delta_{\nu\nu_1'}\delta_{\sigma\sigma_1}\delta_{\bk,\bk_1+\bq} \ee^{\ii[(\varepsilon_{\nu'\bk'}-\mu)(t-t_2)-(\varepsilon_{\nu\bk}-\mu)(t-t_1)]}}
% \end{split}
% \end{equation}
can be evaluated by factorizing the average into low-energy electrons that are inside the system and subject to the system Hamiltonian $\hat{H}$ and high-energy photoemitted electrons subject to a completely single-particle and spin-independent Hamiltonian. We further impose $\bq\approx0$ for small photon momentum, $\bk\simeq\bk',\nu=\nu'$ for sharp momentum distribution of the photoelectrons arrived at the detector, and the energy relation $\varepsilon_{\nu\bk}-\mu=\omega_\bq-\varepsilon$. The result reads
\begin{equation}
\begin{split}
    &P_{\nu_1\sigma_1\nu_2\sigma_2}(\bk)
    =-\ii \sum_{\nu_1\nu_2} \int\dd t_1 \dd t_2 M^*\mycomment{_{\bq=0}}(\nu_2,\sigma_2,\nu,\bk,t_2)M\mycomment{_{\bq=0}}(\nu_1,\sigma_1,\nu,\bk,t_1) s(t_1)s(t_2)\ee^{\ii\varepsilon(t_2-t_1)}G^<_{\nu_1\sigma_1\nu_2\sigma_2}(\bk,t_1,t_2),
\end{split}
\end{equation}
which reduces to the matrix form in the main text as we do not have the $\nu_1,\nu_2$ indices and we also take the featureless matrix element approximation. 

One can prove the physical reality $P_i(\varepsilon,\bk, t) \in \mathds{R}$ given in the main text by casting the intensity matrix $P(\varepsilon,\bk, t)$ into
\begin{equation}
    P(\varepsilon,\bk,t)=  -\frac{\ii}{2\mycomment{\hbar^2}}\int_{-\infty}^{\infty} \dd t_1 \dd t_2\, [C(t_1,t_2)-C^\dag(t_1,t_2)]
\end{equation}
with the manifestly anti-Hermitian integrand and $C(t_1,t_2) = \ee^{\ii\varepsilon (t_1-t_2)} s(t_1-t)s(t_2-t) G^<(\bk,t_1,t_2)$ satisfying $C^\dag(t_2,t_1)=-C(t_1,t_2)$.
A further physical condition is that all diagonal elements 
\begin{equation}
    P_{a\pm,a\pm}=P_0\pm P_a \geq 0 
\end{equation}
along \textit{any} quantization axis ($a=1,2,3$), as physically required by the \textit{positivity} of the photocurrent intensity $I_{aa}$. 
An approximated gauge invariance ansatz of substituting the momentum by $\tilde\bk(t_1,t_2) = \bk + \frac{e}{\hbar}\frac{1}{t_1-t_2}\int_{t_2}^{t_1} \bA(\tau)\dd\tau$ has been proposed, but does not guarantee the positivity for multiband cases\cite{Freericks2015_SI,Freericks2016_SI}.
As we put our focus on times when the pump pulse considerably decays, it suffices to use a specific gauge, e.g., the Hamiltonian gauge we adopt, and this positivity can be naturally confirmed in our calculation.

\section{Keldysh response theory}\label{SM:KeldyshLinear}

\subsection{Analytical expression of removal Green's function}
We push the analytical result Eq.~\eqref{eq:XYZmualpha_1storder} further to perform the time convolution in Eq.~\eqref{eq:KeldyshGra<}. 
From the building-block function Eq.~\eqref{eq:Ierf1} we can define 
\begin{equation}\label{eq:Ia}
    I_{\alpha}(T) = I(\alpha\omega,2d,T),\quad \alpha=\pm
\end{equation}
and
\begin{equation}\label{eq:Bcs}
\begin{split}
    B_{c}=I_{+} + I_{-}, \quad B_{s}=I_{+} - I_{-}.
\end{split}
\end{equation}
The $\ee^{-\ii t_-d_0}$ factor everywhere in Eq.~\eqref{eq:XYZmualpha_1storder} remains.
According to Eq.~\eqref{eq:KeldyshGra<}, we would need the semi-infinite time convolution 
\begin{equation}
    (A*g)(t')=\int_{-\infty}^T A_{c,s}(t)g(t'-t)\dd\tau
\end{equation}
where $t'=(t_1+t_2)/2$,  $A_c(t)=\ee^{-\frac{t^2}{2t_0^2}}\cos{\omega t}$ or $A_s(t)=\ee^{-\frac{t^2}{2t_0^2}}\sin{\omega t}$ is the $t$-dependent part in the vector potential $\bA(t)$, and $g(t)$ ranges among the several (complex) trigonometric functions in Eq.~\eqref{eq:XYZmualpha_1storder} dependent on $ t_\pm$ and even $b=d\beta$.
Direct calculation gives us 
\begin{equation}
\begin{split}
    &A_c * \cos{d t_+} \rightarrow  \mathrm{Re}[B_{c}],\quad
    A_c * \sin{d t_+} \rightarrow  \mathrm{Im}[B_{c}],\\
    &A_s * \cos{d t_+} \rightarrow  \mathrm{Im}[B_{s}],\quad
    A_s * \sin{d t_+} \rightarrow  -\mathrm{Re}[B_{s}],
\end{split}
\end{equation}
where we use the property $I(\alpha\omega,a)^*=I(-\alpha\omega,-a)$.
And similarly, 
\begin{equation}\label{eq:A_cs_Conv_ibeta}
\begin{split}
    &A_c * \cos{(d t_+\pm \ii\beta)} \rightarrow  \mathrm{Re}[B_{c}] \cosh{b} \mp \ii\, \mathrm{Im}[B_{c}]\sinh{b},\quad
    A_c * \sin{(d t_+\pm \ii\beta)} \rightarrow  \mathrm{Im}[B_{c}] \cosh{b} \pm \ii\, \mathrm{Re}[B_{c}]\sinh{b},\\
    &A_s * \cos{(d t_+\pm \ii\beta)} \rightarrow  \mathrm{Im}[B_{s}] \cosh{b} \mp \ii\, (-\mathrm{Re}[B_{s}])\sinh{b},\quad
    A_s * \sin{(d t_+\pm \ii\beta)} \rightarrow  (-\mathrm{Re}[B_{s}]) \cosh{b} \pm \ii\, \mathrm{Im}[B_{s}]\sinh{b}.
\end{split}
\end{equation}
As aforementioned, for the terms with  $\cos{d t_-},\sin{d t_-},\cos{d( t_-\pm\ii\beta)},\sin{d( t_-\pm\ii\beta)}$ in Eq.~\eqref{eq:XYZmualpha_1storder}, effectively we can simply use Eq.~\eqref{eq:A_cs_Conv_ibeta} with $ t_+\rightarrow t_-$ and 
\begin{equation}
\begin{split}
    &A_c * \cos{d t_-} \rightarrow  \mathrm{Re}[B_{c}]\big|_{d=0}\cos{d t_-},\quad
    A_c * \sin{d t_-} \rightarrow  \mathrm{Re}[B_{c}]\big|_{d=0}\sin{d t_-},\\
    &A_s * \cos{d t_-} \rightarrow  \mathrm{Im}[B_{s}]\big|_{d=0}\cos{d t_-},\quad
    A_s * \sin{d t_-} \rightarrow  \mathrm{Im}[B_{s}]\big|_{d=0}\sin{d t_-}
\end{split}
\end{equation}
In one word, all we need for the time-convolution is to evaluate Eq.~\eqref{eq:Ia}, which basically comprises four distinct complex-valued $\mathrm{Erf}$'s $I(\pm\Omega,d(\bk),t_{1,2})$ for given $\Omega,\bk$ and another four $I(\pm\Omega,0,t_{1,2})$ for given $\Omega$. %And this is implemented, e.g., in the \href{http://ab-initio.mit.edu/wiki/index.php/Faddeeva_Package}{Faddeeva Package} with both C and Python interfaces.

\subsection{Analytical expression of late-time SARPES signal}
We redefine $t_+=(t_1+t_2)/2,t_-=t_1-t_2$ as the Wigner-Weyl coordinates in the following. According to Eqs.~\eqref{eq:Ia}\eqref{eq:Bcs}, at late times we can approximate the error functions therein
\begin{equation}
    \mathrm{Erf}(\frac{t_{1,2}-\ii(\pm\Omega- 2d)t_0^2}{\sqrt{2}t_0})|_{t_{1,2}\gg t_0}\approx1
\end{equation}
and then have $B_{c,s}(t_{1,2}) = \ee^{\ii2dt_+} \,\tilde{W}_{c,s}$ that becomes the same for the two parts in Eq.~\eqref{eq:KeldyshGra<} with 
\begin{equation}\label{eq:W_cs}
    \tilde{W}_{c,s} =\sqrt{\frac{\pi}{2}} t_0  \sum_{a=\pm}a^x\,\ee^{-\frac{t_0^2}{2}(a\Omega-2d)^2}
\end{equation}
where $x=0,1$ respectively for $\tilde{W}_{c,s}$. Then we can write Eq.~\eqref{eq:KeldyshGra<} in a concise form
\begin{equation}
\begin{split}
    \mathscr{G}_0(\bk,t_1,t_2)&\equiv 0 \\
    \mathscr{G}_i(\bk,t_1,t_2) &= \frac{2\ii A_0\ee^{-\ii d_0t_-}\sin{d\beta}}{d(\cosh{d\beta}+\cosh{(d_0-\mu_0)\beta})} \left\{ \left[\tau \tilde{W}_s (d\partial_2\bd-\bd\partial_2d) + \tilde{W}_c\bd\times\partial_1\bd\right]\cos{2dt_+} \right.\\
    &\left.+ \left[-\tilde{W}_c (d\partial_1\bd-\bd\partial_1d) + \tau \tilde{W}_s\bd\times\partial_2\bd\right]\sin{2dt_+} \right\}
\end{split}
\end{equation}
with $\partial_\mu=\partial_{k_\mu}$. 
Now we plug this into the SARPES signal formulae in the main text, for which we need a prototype integral
\begin{equation}\label{eq:doubletime_integral}
\begin{split}
    I(\varepsilon,t)&=\int_{-\infty}^{\infty} \int_{-\infty}^{\infty} \dd t_1 \dd t_2\, \ee^{\ii\varepsilon (t_1-t_2)} s(t_1-t)s(t_2-t) \ee^{-\ii d_0t_-}[C \cos{2dt_+} + D \sin{2dt_+}] \\
    &= \frac{1}{2\pi t_\mathrm{pb}^2}\int_{-\infty}^{\infty} \int_{-\infty}^{\infty} \dd t_+ \dd t_-\, \ee^{\ii(\varepsilon-d_0) t_-} \ee^{-\frac{1}{2 t_\mathrm{pb}^2}\left[2(t_+-t)^2+\frac{t_-^2}{2}\right]} [C \cos{2dt_+} + D \sin{2dt_+}] \\
    &= \frac{1}{2\pi t_\mathrm{pb}^2}
    \left(\int_{-\infty}^{\infty}  \dd t_-\, \ee^{\ii(\varepsilon-d_0) t_-} \ee^{-t_-^2/4 t_\mathrm{pb}^2} \right)
    \left(\int_{-\infty}^{\infty} \dd t_+ \ee^{-\frac{(t_+-t)^2}{ t_\mathrm{pb}^2}}[C \cos{2dt_+} + D \sin{2dt_+}] \right) \\
    &= \frac{1}{2\pi t_\mathrm{pb}^2} \times 2\sqrt{\pi}t_\mathrm{pb} \ee^{-(\varepsilon-d_0)^2t_\mathrm{pb}^2} \times \sqrt{\pi} t_\mathrm{pb} \ee^{-d^2t_\mathrm{pb}^2}(C \cos{2dt} + D \sin{2dt}) \\
    &= \bar{F}(\varepsilon) (C \cos{2dt} + D \sin{2dt})
\end{split}
\end{equation}
with 
\begin{equation}\label{eq:energy-dependence0}
    \bar{F}(\varepsilon)=\ee^{-[(\varepsilon-d_0)^2+d^2]t_\mathrm{pb}^2}.
\end{equation}
Using the identity $$\frac{\sinh{d\beta}}{\cosh{d\beta}+\cosh{(d_0-\mu_0)\beta}}=f(\varepsilon_-) - f(\varepsilon_+),$$
we then arrive at the most general form Eq.~\eqref{eq:P_general} of the late-time SARPES signal for a two-band model.
And we are ready to study the tornado topology hidden herein.

\subsection{Momentum envelope function in tornado response}\label{SM:envelope_W_k}
We can rewrite the momentum envelope function in Eq.~\eqref{eq:P_TI_main}
\begin{equation}\label{eq:momentum-dependence}
\begin{split}
    W(\bk)&=\sqrt{\frac{\pi}{2}} t_0 \ee^{-\frac{t_0^2}{2}(\Omega-2d(\bk))^2-d(\bk)^2t_\mathrm{pb}^2}
    =\sqrt{\frac{\pi}{2}} t_0 \ee^{-t_0^2b\left[\left(d(\bk)-\frac{\Omega}{b}\right)^2+\frac{\Omega^2}{2b}\left(1-\frac{2}{b}\right)\right]},
\end{split}
\end{equation}
where $b=2+\left(\frac{t_\mathrm{pb}}{t_0}\right)^2$. %Note that one can obviously as well keep $W_{c,s}$ and reach essentially the same conclusions since they are just factors purely dependent on $d$ rather than $\bd$; on the other hand, the most important parts are the terms following $W_{c,s}$. 
Eq.~\eqref{eq:momentum-dependence} gives the $d(\bk)$- and hence $\bk$-dependence of the signal. It bears a peak ring or annulus at $vk_0=\sqrt{\Omega^2/b^2-m^2}$ when $bm<\Omega$ and only one maximum at the origin $k=0$ otherwise. Practically, in order to observe considerable signal strength even inside the ring, we can require 
\begin{equation}
    W(k=0)>\ee^{-\xi}W(k_0),
\end{equation}
e.g., for $\xi=1$, which gives $\Omega>bm>\Omega-\sqrt{\xi b/t_0^2}$. Therefore, we have two cases of the $k$-dependence of the signal
\begin{itemize}
    \item Disk-like tornado signal when $bm>\Omega-\sqrt{\xi b/t_0^2}$, e.g., $\xi=1$. This is typically the case when we have big enough $m,t_\mathrm{pb}$ and/or small enough $\Omega,t_0$. The expansion reads $W(0+\delta k) \approx W(k=0)  \ee^{-t_0^2bv^2(1-\frac{\Omega}{bm})\delta k^2}$ when $bm>\Omega$. In fact, two simple and useful conclusions in this case are 
    \begin{itemize}
        \item $W(k=0)$ maximizes at $\Omega=\Delta=2m$ (note that $b>2$ and hence peak at origin always holds);
        \item Lowering $t_0$ gives larger and less annulus-like signal; lowering $t_\mathrm{pb}$  gives larger but more annulus-like signal. 
        To get larger and more center-peaked (i.e., less annulus-like) signal there are two ways: smaller $t_0$; smaller $t_\mathrm{pb}$ while fixing $b$, i.e., $\frac{t_\mathrm{pb}}{t_0}$.
    \end{itemize}
    \item Annulus-like tornado signal otherwise. The expansion reads $W(k_0+\delta k) \approx W(k_0) \ee^{-t_0^2bv^2(1-\frac{b^2m^2}{\Omega^2})\delta k^2}$.
\end{itemize}

\section{Topological tornado response}\label{SM:topo}
Here we present the full theory accounting for the topological tornado responses, which is based on Eq.~\eqref{eq:P_TI_main1}. We set $v=1$ for simplicity and put most appearances of the three helicity factors in color as $\blue{\chi},\red{\tau},\green{\nu}$ in order to facilitate identification.

\subsection{Out-of-plane \texorpdfstring{$z$}{z}-component}
Let's quickly check the $\tilde{P}_3$ response of $p$-wave-like form, which certainly falls in Eq.~\eqref{eq:tornado_form}
\begin{equation}\label{eq:z_winding}
\begin{split}
    \tilde{P}_3(\bk,t)
    &=\sqrt{k_x^2(m+d\blue{\chi}\red{\tau})^2+k_y^2(\blue{\chi}d+m\red{\tau})^2} \sin{[2dt-\arctan(k_x(m+\blue{\chi}\red{\tau}d),k_y(\blue{\chi}d+\red{\tau}m))]}\\
    &=
    \begin{cases}
    k d_{\blue{\chi}\red{\tau}}\sin{[2dt+\frac{\pi}{2}-\red{\tau}(\theta_\bk+\blue{\chi}\frac{\pi}{2})]} & \red{\tau}=\pm1 \\
    \sqrt{m^2k_x^2+d^2k_y^2} \sin{[2dt+\frac{\pi}{2}-\green{\nu}(\blue{\chi}\arctan(|m|k_x,dk_y)+\frac{\pi}{2})]} & \red{\tau}=0
    \end{cases},
\end{split}
\end{equation}
where we denote $d_\pm=d \pm m$.
We clearly see $\tilde{P}_3(\red{\tau}=0)$ corroborates with the $\delta$-pulse calculation with $\Xi=\blue{\chi}\green{\nu}$. Also, $\tilde{P}_3(\red{\tau}=\pm1)$ readily shows the helicity driven by the extrinsic $\red{\tau}$, giving rise to $\Xi=\red{\tau}$ as expected, summarized in 
\begin{equation}\label{eq:Xi-chi_result}
\begin{split}
    W_s=
    \begin{cases}
    \red{\tau} & \red{\tau}=\pm1 \\
    \blue{\chi}\green{\nu} & \red{\tau}=0
    \end{cases}
\end{split}
\end{equation}
for the surface state with an intrinsic helicity $\blue{\chi}$ and sign of mass $\green{\nu}$.
Besides, the gapless case $m=0$ obviously only renders the tornado in the $\red{\tau}=0$ case absent since $\Theta(\theta_\bk)=\pm\pi/2$.

We also note that the prefactor $d_{\blue{\chi}\red{\tau}}$ in Eq.~\eqref{eq:z_winding} explains the strong or weak dichroic response strength.

\subsection{In-plane amplitude}\label{Sec:inplane_amplitude}
For the in-plane spin texture concerning $\tilde{P}_{1,2}$, it is not as transparent as the $\tilde{P}_3$ case. Let's consider the most relevant 2D vector field $\tilde{\bP}_\mathrm{in}$, henceforth denoted as $\bw$ for notational brevity
\begin{equation}\label{eq:inplaneW}
    \bw(\bk)  =-\left(  (dm+\blue{\chi}\red{\tau}(d^2-k_y^2))\cos{2dt} + \blue{\chi}k_xk_y \sin{2dt}, \; \red{\tau} k_xk_y\cos{2dt} + (\blue{\chi}\red{\tau} dm+d^2-k_x^2) \sin{2dt} \right).
\end{equation}

Firstly, for its amplitude, we have the $d$-wave-like expression instead of the $p$-wave-like $\tilde{P}_3$
\begin{equation}\label{eq:w^2_1}
\begin{split}
    w^2(\bk)=|\bw(\bk)|^2=    
    \begin{cases}
    \frac{1}{2}d_{\blue{\chi}\red{\tau}}^2 \left[ (d^2+m^2) + (k_x^2-k_y^2)\cos{4dt} +\red{\tau} 2k_xk_y \sin{4dt} \right] & \red{\tau}=\pm1 \\
    \frac{1}{2}\left\{ (d^2+m^2)(d^2-k_x^2) + (m^2k_x^2-d^2k_y^2)\cos{4dt} + \blue{\chi}2mdk_xk_y \sin{4dt} \right\} & \red{\tau}=0
    \end{cases}.
\end{split}
\end{equation}
We readily see that the time-dependent part of $w^2(\bk)$ reads 
\begin{equation}\label{eq:w^2_2}
\begin{split}
    \begin{cases}
    \frac{1}{2}d_{\blue{\chi}\red{\tau}}^2 k^2\sin{(4dt+\frac{\pi}{2} -\red{\tau} 2\theta_\bk)} & \red{\tau}=\pm1 \\
    \frac{k^2}{4}(D_+^2-D_-^2\cos{2\theta_\bk}) \sin{\left[ 4dt +\pi-(\blue{\chi}\green{\nu}\arctan{(\frac{1}{2} (D_+^2\cos{2\theta_\bk}-D_-^2),|m|d\sin{2\theta_\bk})} +\green{\nu}\frac{\pi}{2})\right]} & \red{\tau}=0
    \end{cases}
\end{split}
\end{equation}
where we denote $D_\pm^2=d^2\pm m^2$, which again falls in Eq.~\eqref{eq:tornado_form}. The $\red{\tau}=0$ case follows the intrinsic chirality and the complexity disappears if we approximately set $d=m$ in the coefficients, which simply gives
\begin{equation}
    \frac{k^2}{2}m^2 \sin{(4dt+\frac{\pi}{2}-2\theta_\bk)}.
\end{equation}
Besides, the gapless case $m=0$ obviously only renders the tornado in the $\red{\tau}=0$ case absent since $\Theta(\theta_\bk)=\pi$.

We also note that Eq.~\eqref{eq:w^2_2} essentially follows all the topological features of Eq.~\eqref{eq:z_winding}.

\subsection{In-plane angle winding}\label{Sec:inplane_winding}
Secondly, let's look at the information involving the azimuthal angle $\phi$ of $\bw(\bk)$. 
\begin{itemize}
    \item We rewrite $-\bw=\bu + \bv$
where 
\begin{equation}\label{eq:u&v_split}
    \bu = \left(\bk_{\red{\tau}}\cdot\hat{\bq}\right)\bk_{\blue{\chi}},\quad \bv = m\begin{pmatrix} d+{\blue{\chi}\red{\tau}} m & \\ & {\blue{\chi}\red{\tau}} d+m \end{pmatrix}
    \hat{\bq}
\end{equation}
with $\bk_\pm=(\pm k_x,k_y),\hat{\bq} = (\cos{2dt},\sin{2dt})$. Given $k$, i.e., a circle $C_k$ on the 2D $\bk$-plane, $\bv$ is a constant vector field. On the other hand, while $\bu$ is oriented parallel to the radial direction of $\hat{\bk}_{\blue{\chi}}$ it vanishes at two diametrically opposite points on $C_k$ where $\bk_{\red{\tau}}\perp\hat{\bq}$. In fact, the vector field $\bu=(u_x,u_y)$ maps $C_k$ to a new trajectory, a \textit{circle} $\mathcal{C}_k$ that is \textit{doubly and $\blue{\chi}$-clockwisely traversed} and also \textit{passes the origin} twice. This can be seen in polar coordinates 
\begin{equation}\label{eq:u_trajectory}
    \bu = k^2({\red{\tau}} \hat{q}_x\cos\theta_\bk+\hat{q}_y\sin\theta_\bk)(\blue{\chi}\cos\theta_\bk,\sin\theta_\bk), 
\end{equation}
which leads to the parametric equation of $\bu$'s trajectory in the form of a circle that crosses the origin
\begin{equation}
    (u_x-{\blue{\chi}\red{\tau}}\frac{k^2}{2}\hat{q}_x)^2 + (u_y-\frac{k^2}{2}\hat{q}_y)^2 = ({\red{\tau}}^2\hat{q}_x^2+\hat{q}_y^2)\frac{k^4}{4}.
\end{equation}
Too see this, we can denote $R=k^2({\red{\tau}} \hat{q}_x\cos\theta_\bk+\hat{q}_y\sin\theta_\bk)$, leading to $u_x^2+u_y^2=R^2=Rk^2({\red{\tau}} \hat{q}_x\cos\theta_\bk+\hat{q}_y\sin\theta_\bk)=k^2({\blue{\chi}\red{\tau}} \hat{q}_xu_x+\hat{q}_yu_y)$.
Since $\bv$ is a constant vector along $C_k$, adding $\bv$ to $\bu$, i.e., $-\bw$, simply translates $\bu$'s trajectory circle $\mathcal{C}_k$ to a new circle $\mathscr{C}_k$ with its origin at
\begin{equation}
    -\bw_0=\left[\frac{k^2}{2}\begin{pmatrix} {\blue{\chi}\red{\tau}} & \\ & 1 \end{pmatrix}
    +m\begin{pmatrix} d+{\blue{\chi}\red{\tau}} m & \\ & {\blue{\chi}\red{\tau}} d+m \end{pmatrix}\right]
    \hat{\bq}.
\end{equation}
We define 
\begin{equation}
    f_{\blue{\chi},\red{\tau}}(\bw)= (\bw-\bw_0)^2 - ({\red{\tau}}^2\hat{q}_x^2+\hat{q}_y^2)\frac{k^4}{4}
\end{equation}
and have 
\begin{equation}
f_{\blue{\chi},\red{\tau}}(\bzero)=
    \begin{cases}
    {\blue{\chi}\red{\tau}} m (d+{\blue{\chi}\red{\tau}} m) \left(k^2+{\blue{\chi}\red{\tau}} m (d+{\blue{\chi}\red{\tau}} m)\right) & {\blue{\chi}\red{\tau}}=\pm1 \\
    m^2 d^2  & {\red{\tau}}=0
    \end{cases}.
\end{equation}
A key observation is that 
\begin{equation}
    \begin{cases}
    \green{\nu}\,\blue{\chi}\,{\red{\tau}} \,f_{\blue{\chi},{\red{\tau}}}(\bzero) >0 & {\blue{\chi}\red{\tau}}=\pm1 \\
    f_{\blue{\chi},{\red{\tau}}}(\bzero)>0 & {\red{\tau}}=0
    \end{cases},
\end{equation}
where the inequalities hold as long as $m\neq0,k>0$.
This leads to
\begin{equation}
    \begin{cases}
    \textrm{$\bw=\bzero$ lies outside $\mathscr{C}_k$} & \textrm{${\red{\tau}}=0$ or ${\blue{\chi}\red{\tau}}\green{\nu}=1$} \\
    \textrm{$\bw=\bzero$ lies inside $\mathscr{C}_k$} & {\blue{\chi}\red{\tau}}\green{\nu}=-1
    \end{cases},
\end{equation}
which immediately dictates the winding number (note that $\bu$ and $\bw$ share the same revolving sense)
\begin{equation}\label{eq:xy_winding}
    w_\phi=\int_0^{2\pi}\dd\theta_\bk\arctan(w_x,w_y)=\begin{cases}
    0 & \textrm{${\red{\tau}}=0$ or ${\blue{\chi}\red{\tau}}\green{\nu}=1$} \\
    2\blue{\chi} & {\blue{\chi}\red{\tau}}\green{\nu}=-1
    \end{cases}.
\end{equation}

As $k$ grows, the rotation of $\hat{\bq}$ or $\bv$, together with the directly related rotation of $\bu$ seen from its origin $\bw_0(k)$, is possible to generate the spiral structure. This, however, depends on whether $\bw$ can trace all the directions. Therefore, such a $w_\phi=2\blue{\chi}$ winding exactly accounts for the appearance of two spiral arms that we only see in plotting $\phi=\arctan(w_x,w_y)$ for the following four cases: $m>0,\blue{\chi}=\pm1,\red{\tau}=\mp1$ and $m<0,\blue{\chi}=\pm1,\red{\tau}=\pm1$.

\item For the gapless or nearly gapless cases, i.e., when $|m|t<1$, the situation of $\phi$ is different. While the orientation (color) rotation sense still follows the exact Eq.~\eqref{eq:xy_winding}, which becomes ill-defined (i.e., not fully winding around but rotation sense still discernible) only when $m=0$, we also have an envelope spiral shape clearer and clearer with decreasing $mt$
\begin{equation}
    \begin{cases}
    \textrm{spiral of helicity ${\red{\tau}}$} & {\red{\tau}}=\pm1 \\
    \textrm{no spiral when $m=0$; otherwise spiral of helicity $\blue{\chi}\green{\nu}$} & {\red{\tau}}=0
    \end{cases}.
\end{equation}
While the crossover regime $|m|t\sim1$ can be a complex smooth connection between the two cases, it is beneficial to see the $m=0$ case. Now since $\bv=0$, we only have $\bu$ from Eq.~\eqref{eq:u&v_split}. As $k$ or $d$ grows the unit vector $\hat{\bq}$ rotates, the corresponding variation of $\bu$ in the prefactor 
\begin{equation}\label{eq:tornado_m=0_prefactor}
    \bk_{\red{\tau}}\cdot\hat{\bq}=
    \begin{cases}
    k\sin{[2dt-{\red{\tau}}(\theta_\bk-\frac{\pi}{2})]} & {\red{\tau}}=\pm1 \\
    k\sin{2dt}\sin{\theta_\bk} & {\red{\tau}}=0
    \end{cases}
\end{equation} 
can be compensated by an appropriate rotation in $\theta_\bk$ as long as ${\red{\tau}}=\pm1$, simply because $\bk_{{\red{\tau}}=0}$ has a fixed direction only. 

This serves as the origin of the spiral shape formation in the externally driven ${\red{\tau}}=\pm1$ cases. Obviously, this falls in Eq.~\eqref{eq:tornado_form} and we have the winding number $W_s={\red{\tau}}$ for ${\red{\tau}}=\pm1$. The reason why there are two instead of only one arms is that the function plotted is $\arctan\bu$ rather than Eq.~\eqref{eq:tornado_m=0_prefactor}. Note that $\bu(\bk)=\bu(-\bk)$ while $(\bk_{\red{\tau}}\cdot\hat{\bq})\big|_{\bk}=-(\bk_{\red{\tau}}\cdot\hat{\bq})\big|_{-\bk}$, 
which implies that while $\bk_{\red{\tau}}\cdot\hat{\bq}$ has 1 positive and 1 negative arms (i.e., $\mathcal{R}_s=1$ repeating arm) $\arctan\bu$ has two repeating arms.
Actually, the envelope of finite $\bu$ vector field is bounded by the contour curve of $\bk_{\red{\tau}}\cdot\hat{\bq}=0$, which evidently gives rise to spiral only for ${\red{\tau}}=\pm1$. In fact, the trajectory of $\bk_{\red{\tau}}\cdot\hat{\bq}=0$ is simply given in polar coordinates 
\begin{equation}
    \theta_\bk=2kt\pm\frac{\pi}{2},
\end{equation}
i.e., two Archimedean spirals for the two repeating arms, at which $\arctan{(\bw=\bzero)}$ exhibits a singular $\pi$-jump. This $\pi$-jump spiral is also why the radial correpondence $\sgn{(\partial_k\phi)=\sgn{(m)}}$ becomes ill-defined. Now, as $\bu$ itself always passes through the origin, which exactly corresponds the this $\pi$-jump, its winding can only complete a half and hence the absence of the massive topological winding of $\phi$. But still, in the incomplete winding, the variation or rotation sense of $\phi$ follows the same helicity $\blue{\chi}$ as the massive case.
\end{itemize}

\subsection{Radial correspondence}\label{Sec:radial_correspondence}
Let's lastly inspect the robust correspondence $\mathcal{K}=\sgn{(\partial_k\phi)}=\sgn{(m)}$ in the in-plane signal $\phi(\bk)=\arctan{\bw}$ of Eq.~\eqref{eq:inplaneW}. Using the one-form $\dd\phi = \frac{1}{w^2}(w_x \dd w_y -w_y \dd w_x)$ that is continuous everywhere except at the origin $\bw=0$, we have 
\begin{equation}
    \partial_k\phi = \frac{1}{w^2}(w_x \partial_k w_y -w_y \partial_k w_x).
\end{equation}
We hence need to study the positivity of 
\begin{equation}\label{eq:radial_positivity}
\begin{split}
    K&=\frac{2d}{km}(w_x \partial_k w_y -w_y \partial_k w_x) \\
    &=\begin{cases}
    \left[4dt\left(2d\,d_{\blue{\chi}\red{\tau}}-k^2\right)-d_{\blue{\chi}\red{\tau}}^2\sin{(\red{\tau}2{\theta_\bk}-4dt)}\right] & \red{\tau}=\pm1 \\
    4dt(k^2\sin^2{{\theta_\bk}}+m^2)-\sin{4dt}[(d^2+m^2)\cos{2{\theta_\bk}}-k^2]/2-\blue{\chi}md(1-\cos{4dt})\sin{2{\theta_\bk}} & \red{\tau}=0
    \end{cases}
\end{split}
\end{equation}
For the $\red{\tau}=\pm1$ case, $K$ is not always positive. But we can show it is positive as long as $t$ is large enough, which is obvious since $\left(2d\,d_{\blue{\chi}\red{\tau}}-k^2\right)>0$ holds when $k,m\neq0$.
In fact, we have the infimum $\underline{K}(t)=\inf_{\theta_\bk}{[K]}=4dt\left(2d\,d_{\blue{\chi}\red{\tau}}-k^2\right)-d_{\blue{\chi}\red{\tau}}^2$ and, for instance, we can prove $\underline{K}(|2m|^{-1})>0$, which gives a safe bound $2|m|t>1$ to ensure $K>0$. 
This can be seen as follows 
\begin{equation}
\begin{split}
    |m|\underline{K}(|2m|^{-1})
    &=[4d^2-|m|(d\pm m)](d\pm m)-2dk^2 \\
    & > (3d^2\mp m^2)(d\pm m)-2dk^2=k^2[3d^2\mp m^2-2d(d\mp m)]\\
    &=k^2(d^2\mp m^2\pm2dm)>0.
\end{split}
\end{equation} 
For the $\red{\tau}=0$ case, we have
\begin{equation}\label{eq:radial_positivity_tau=0}
\begin{split}
    K&=4dt(k^2\sin^2{{\theta_\bk}}+m^2)-\blue{\chi}md\sin{2{\theta_\bk}}-\sin{4dt}[(d^2+m^2)\cos{2{\theta_\bk}}-k^2]/2+\blue{\chi}md\cos{4dt}\sin{2{\theta_\bk}}\\
    &=4dt(k^2\sin^2{{\theta_\bk}}+m^2)-\blue{\chi}md\sin{2{\theta_\bk}}-(k^2\sin^2{{\theta_\bk}}+m^2)\sin{(4dt+\Phi_{\blue{\chi}})}
\end{split}
\end{equation}
and its infimum
\begin{equation}
\begin{split}
    \underline{K}&=\inf_{\theta_\bk}{[K]}=(4dt-1)(k^2\sin^2{{\theta_\bk}}+m^2)-|m|d 
\end{split}
\end{equation}
We have $\frac{\partial \underline{K}}{\partial\sin^2{{\theta_\bk}}}=(4dt-1)k^2>0$ as long as $4dt>1$, which can be satisfied by taking $t>\frac{1}{4|m|}$. Under this condition, we have $K>\underline{K}({\theta_\bk}=0)=4m^2dt-|m|(d+|m|)>0$ as long as $t>\frac{1}{2|m|}$.

In summary, we have 
\begin{equation}
    \sgn{(\partial_k\phi)}=\green{\nu}
\end{equation} 
when $t>\frac{1}{2|m|}$ \textit{regardless} of $\blue{\chi}$ and $\red{\tau}$.

\section{\texorpdfstring{$\delta$}{delta}-pulse for LP light}\label{SM:delta}
Here we give the full expression of the SARPES signal under an LP light $\delta$-pulse
\begin{equation}\label{eq:rho_SARPES_pulse}
\begin{split}
    P_0(\varepsilon,\bk,t)&=P_0^{(0)}(\varepsilon,\bk)+\frac{4\alpha\left(f_{\varepsilon_-} - f_{\varepsilon_+}\right)}{(1+\alpha ^2)^2d^3}d\,E_+ \\
    \bP(\varepsilon,\bk,t)&=\bP^{(0)}(\varepsilon,\bk)+\frac{4\alpha\left(f_{\varepsilon_-} - f_{\varepsilon_+}\right)}{(1+\alpha ^2)^2d^3}\left[\bd E_- + \bar{F}(\varepsilon)\bZ(t)\right]
\end{split}
\end{equation}
where 
\begin{equation}
\begin{split}
    \bZ(t)=\begin{pmatrix}
    [-(1-\alpha^2)md^2-2\blue{\chi}\alpha k_x^2k_y]\cos{2dt} +dk_x[2\alpha m-\blue{\chi}(1-\alpha ^2)k_y]\sin{2dt} \\
    (d^2-k_x^2)[-2\alpha k_x\cos{2dt} -d(1-\alpha ^2)\sin{2dt}] \\
    [-\blue{\chi}(1-\alpha^2)d^2k_y+2\alpha mk_x^2]\cos{2dt} +dk_x[2\blue{\chi}\alpha k_y+(1-\alpha ^2)m]\sin{2dt} 
    \end{pmatrix}
\end{split}
\end{equation}
and the equilibrium SARPES signal
\begin{equation}\label{eq:rho0_SARPES_pulse}
\begin{split}
    P_0^{(0)}(\varepsilon,\bk)
    &= f_{\varepsilon_+}\, F_+(\varepsilon) + f_{\varepsilon_-}\, F_-(\varepsilon)  \\
    \bP^{(0)}(\varepsilon,\bk) 
    &= \frac{\bd}{d} \left[f_{\varepsilon_+}\, F_+(\varepsilon) - f_{\varepsilon_-}\, F_-(\varepsilon)  \right].
\end{split}
\end{equation}
Other quantity definitions are already given in the main text. One can observe several properties from Eq.~\eqref{eq:rho_SARPES_pulse}
\begin{itemize}
    \item A salient feature is that the second part in the spin channel contributes the \textit{only} time-dependent signal
\begin{equation}\label{eq:rho_t_SARPES_pulse}
    \bP'(\varepsilon,\bk,t)=\frac{4\alpha}{(1+\alpha ^2)^2d^3} \left(f_{\varepsilon_-} - f_{\varepsilon_+}\right) \bar{F}(\varepsilon)\bZ(t),
\end{equation}
which bears the common energy profile $\bar{F}(\varepsilon)$ as the linear response result.
\item Only this time-dependent $\bP'(\varepsilon,\bk,t)$ has $\alpha$-odd (including the linear response) contributions while all others are $\alpha$-even.
\item Terms proportional to $\left(f_{\varepsilon_-} - f_{\varepsilon_+}\right)$ are crucial to contribute to either the time-dependent (due to virtual excitations) or the time-independent (due to real excitations) deviation away from equilibrium, which plausibly manifests the optical inertness of both two bands being empty or filled.
\item Taking $P_0(\varepsilon,\bk,t)$ as an example, the factor $\left[ F_+(\varepsilon) - F_-(\varepsilon)  \right]$ in $E_+$ exactly relates to the real pumping from the lower $\varepsilon_-$ band to the higher unoccupied $\varepsilon_+$ band. Besides, when $k_y=m=0$ we have $E_+=0$, i.e., there is no real transition, which is because in this case the pumping interaction commutes with $H_0$.
%\item The equilibrium value Eq.~\eqref{eq:rho0_SARPES_pulse} enters through the $c_{12}$ term.
\end{itemize}

\subsection{Match with linear response}
We can use Eq.~\eqref{eq:rho_SARPES_pulse} to obtain the leading photoinduced part
\begin{equation}\label{eq:rho_pulse_SARPES_linear}
\begin{split}
    P_0^{(1)}(\varepsilon,\bk,t)=0 ,\qquad
    P_j^{(1)}(\varepsilon,\bk,t)=\frac{4\alpha}{d^2}\left(f(\varepsilon_-) - f(\varepsilon_+)\right)\bar{F}(\varepsilon)Z_j^{(0)}(t).
\end{split}
\end{equation}
As a sanity check, let's take the zero-temperature limit, leading immediately to 
\begin{equation}
\begin{split}
    P_j^{(1)}(\varepsilon,\bk,t)\big|_{\beta\rightarrow\infty} = \theta(d-|d_0-\mu_0|) \frac{4\alpha}{d^2}\bar{F}(\varepsilon)Z_j^{(0)}(t),
\end{split}
\end{equation}
where the step function $\theta(d-|d_0-\mu_0|)$ appears since any finite response, even due to virtual excitations captured by the leading-order response, requires at least finite occupation in the lower band.
Most importantly, we find that the linear-response result Eq.~\eqref{eq:P_TI_main} perfectly matches the $\delta$-pulse result Eq.~\eqref{eq:rho_pulse_SARPES_linear} when $\red{\tau}=0$ as it should do, as long as we notice that $W_{c}\rightarrow2\sqrt{\frac{\pi}{2}}t_0,W_{s}\rightarrow0$ from Eq.~\eqref{eq:W_cs} and $2A_0\sqrt{\frac{\pi}{2}}t_0\rightarrow\tilde{A}_0$ when $t_0\rightarrow0$ and set $v=e=\hbar=1$. Here, to fulfill the perfect match, one should use Eq.~\eqref{eq:W_cs} instead of the further approximated $W$ and note that the $\red{\tau}=0$ case does not involve $B_s$ and hence $W_s$. Also, the relation between $A_0$ and $\tilde{A}_0$ is simply fixed by equating $\int_{-\infty}^\infty\dd t A_0 \ee^{-\frac{t}{2t_0^2}}=A_0\sqrt{2\pi}t_0$ and $\int_{-\infty}^\infty\dd t \tilde{A}_0\delta(t)=\tilde{A}_0$ when $t_0\rightarrow0$. This finally gives the correspondence $4\alpha\leftrightarrow2A_0W_c$ that makes the two results identical.  

% For instance, let's consider the time-dependent part when $\alpha\ll 1$
% \begin{equation}\label{eq:pulse_Zt}
% \begin{split}
%     Z_3(t)/d&\approx -dk_y \cos{2dt} +mk_x\sin{2dt} \\
%     &= \sqrt{m^2k_x^2+d^2k_y^2} \sin{[2dt-\arctan(mk_x,dk_y)]},
% \end{split}
% \end{equation}
% which is of the same form as Eq.~\eqref{eq:tornado_form} and follows the intrinsic chirality of $H_0(\bk)$, i.e., $\Xi=1$. In a cruder approximation, we set $d=m$ for small momenta in the two coefficients
% \begin{equation}
% \begin{split}
%     Z_3(t)/d^2 &\approx -k_y \cos{2dt} +k_x\sin{2dt} \\
%     & = k \sin{[2d(\bk)t-\theta_\bk]},
% \end{split}
% \end{equation}
% the tornado form becomes even more evident.

\subsection{Nonlinear tornado features}
We then study the topology hidden in this time-dependent nonlinear response Eq.~\eqref{eq:rho_t_SARPES_pulse}, for which we can simply look at $\bZ(t)$.
\paragraph{Out-of-plane \texorpdfstring{$z$}{z}-component}
For the normal direction, we have in the form of Eq.~\eqref{eq:tornado_form}
\begin{equation}
\begin{split}
    Z_3(t) = \sqrt{\left((1-\alpha^2)^2d^2+4\alpha^2k_x^2\right)(k_x^2m^2+k_y^2d^2)} \sin{[2dt+\frac{\pi}{2}-\Theta_\bk]}
\end{split}
\end{equation}
with 
\begin{equation}
    \Theta_\bk=\arctan{\left[-\blue{\chi}(1-\alpha^2)d^2k_y+2\alpha mk_x^2,dk_x\left(2\blue{\chi}\alpha k_y+(1-\alpha ^2)m\right)\right]}.
\end{equation}
The behevior of $\Theta_\bk$ can be seen from three limits
\begin{equation}
\begin{split}
    \Theta_\bk=
    \begin{cases}
    \arctan{\left[d(-\blue{\chi}dk_y,mk_x)\right]}=\green{\nu}(\blue{\chi}\arctan(|m|k_x,dk_y)+\frac{\pi}{2})] & \alpha\ll1 \\
    \arctan{\left[-d\alpha^2(-\blue{\chi}dk_y,mk_x)\right]}=\green{\nu}(\blue{\chi}\arctan(|m|k_x,dk_y)-\frac{\pi}{2})] & \alpha\gg1 \\
    \arctan{\left[2\alpha k_x(mk_x,\blue{\chi}dk_y)\right]} & \alpha\approx1
    \end{cases}.
\end{split}
\end{equation}
Therefore, the tornado helicity $\Xi=\green{\nu}\blue{\chi}$ does \textit{not} change with $\alpha$, except a $\pi$-jump of rotation angle offset at $\alpha=1$. Although $\Theta_\bk$ is not necessarily monotonic with respect to $\theta_\bk$, in general, as long as $\alpha\neq1$ one can see the winding
\begin{equation}
    W=\frac{1}{2\pi}\int_0^{2\pi} \dd\theta_\bk \Theta_\bk = \green{\nu}\blue{\chi}.
\end{equation}

\paragraph{In-plane}
As expected, in this LP light case, the azimuthal angle of $\bZ_\mathrm{in}=(Z_1,Z_2)$ does not exhibit any tornado due to the topological switching described in \ref{Sec:inplane_winding}. This is not altered even with nonlinearity taken into account. We therefore merely look at the amplitude in a similar manner as in \ref{Sec:inplane_amplitude}. 
\begin{equation}
\begin{split}
    Z_\mathrm{in}^2(\bk)=|\bZ_\mathrm{in}(\bk)|^2&=
    D_0 + D_1\cos{4dt}  + D_2\sin{4dt}\\
    D_0=\frac{1}{2}c_+(d^2+m^2)(d^2-k_x^2), & \qquad
    D_1=\bu\cdot\bv, \qquad
    D_2=(\bu\times\bv)_z
\end{split}
\end{equation}
where $\bu=(c_-,c_0),\bv=(c_1,c_2)$ with $c_\pm= (\alpha^2-1)^2d^2 \pm 4\alpha^2k_x^2,c_0=4\alpha(1-\alpha^2)dk_x,c_1=(m^2k_x^2-d^2k_y^2)/2,c_2=\blue{\chi}dmk_xk_y$. This leads to
\begin{equation}
\begin{split}
    Z_\mathrm{in}^2(\bk)&=D_0+\sqrt{\bu^2\bv^2}\sin{\left[4dt+\frac{\pi}{2}-\Theta_\bk\right]} \\
    &= D_0 + \frac{c_+}{2}(m^2k_x^2+d^2k_y^2)\sin{\left[4dt+\frac{\pi}{2}-\Theta_\bk\right]}
\end{split}
\end{equation}
in the form of Eq.~\eqref{eq:tornado_form}.
The behevior of $\Theta_\bk$ can be seen from three limits
\begin{equation}
\begin{split}
    \Theta_\bk=\arctan{(D_1,D_2)}=
    \begin{cases}
    \arctan{\left[d^2(c_1,c_2)\right]} & \alpha\ll1 \\
    \arctan{\left[\alpha^4d^2(c_1,c_2)\right]} & \alpha\gg1 \\
    \arctan{\left[-4\alpha^2 k_x^2(c_1,c_2)\right]} & \alpha\approx1
    \end{cases}.
\end{split}
\end{equation}
Therefore, the tornado helicity $\Xi=\green{\nu}\blue{\chi}$ does \textit{not} change with $\alpha$ except distortion near $\alpha=1$. Although $\Theta_\bk$ is generally not monotonic with respect to $\theta_\bk$, in general, one can see the winding
\begin{equation}
    W=\frac{1}{2\pi}\int_0^{2\pi} \dd\theta_\bk \Theta_\bk = 2\green{\nu}\blue{\chi}.
\end{equation}
We further check the radial correspondence following \ref{Sec:radial_correspondence}, for which we define
\begin{equation}
\begin{split}
    K&=\frac{2}{dkm}(Z_1 \partial_k Z_2 -Z_2 \partial_k Z_1) 
    =
    \begin{cases}
    I_- & \alpha\ll1 \\
    \alpha^4I_- & \alpha\gg1 \\
    \frac{4k^2}{d^2}\cos^2{\theta_\bk}\,I_+ & \alpha\approx1
    \end{cases}
\end{split}
\end{equation}
with $I_\pm=4dt(k^2\sin^2{{\theta_\bk}}+m^2) \pm \sin{4dt}[(d^2+m^2)\cos{2{\theta_\bk}}-k^2]/2-\blue{\chi}md(1 \pm \cos{4dt})\sin{2{\theta_\bk}}$. Following Eq.~\eqref{eq:radial_positivity_tau=0}, we can prove $I_\pm\geq0$ as long as $t>\frac{1}{2|m|}$.
Therefore, also confirmed numerically, we have when $t>\frac{1}{2|m|}$ \textit{regardless} of $\blue{\chi}$ and $\alpha$
\begin{equation}
    \sgn{(\partial_k\phi)}=\green{\nu}.
\end{equation} 

\section{Scale estimation}\label{SM:scale}
Here we estimate the realistic pump field strength as a dimensionless quantity
\begin{equation}
    \gamma=evA_0/\Omega.
\end{equation}
Note that this definition is sensible as it relates to the $\delta$-pulse dimensionless quantity via $\gamma=\alpha/\pi$ when we use the natual identification $\tilde{A}_0=A_0T_0$. The vector potential strength is estimated from 
\begin{equation}
    A_0=E_0/\Omega.
\end{equation}
%There are, however, two options for $T_0$ or $\varepsilon_0$.
% \begin{itemize}
%     \item $T_0=2\pi/\Omega$ given by the pump field frequency; for an ultrashort pulse the pulse width can be $2t_0\sim2\pi/\Omega$.
%     \item $T_0=h/\Delta$ given by the mass gap for the gapful case.
% \end{itemize}
% The smaller $T_0$ (or the larger $\varepsilon_0$) will dominate and determine $\gamma$. Therefore, the gapless case will solely have $T_0=2\pi/\Omega$. On the other hand, when the gap $\Delta\gtrsim\hbar\Omega$ the gap size can be more important.
The electric field strength $E_0$ is directly given as $E_0\sim2.4\times10^5\mathrm{V/m}$\cite{Reimann2018_SI} with THz pump around 1THz, i.e., small $\hbar\Omega\sim4\mathrm{meV}$. Alternatively, we can use the formula for energy flux density $I_0=\frac{c\epsilon_0}{2}E_0^2$. We have, e.g., $I_0\sim\frac{0.05\mathrm{mJ/{cm}^2}}{(3.6\mathrm{MHz})^{-1}}$ with pump fluence $0.05\mathrm{mJ/{cm}^2}$ and repetition rate $3.6\mathrm{MHz}$\cite{Jozwiak2016_SI} and $I_0\sim\frac{0.5\mathrm{mJ/{cm}^2}}{(0.25\mathrm{MHz})^{-1}}$ with pump fluence $0.5\mathrm{mJ/{cm}^2}$ and repetition rate $0.25\mathrm{MHz}$\cite{Cacho2015_SI}, leading respectively to $E_0\sim3.7\times10^4\mathrm{V/m}$ and $E_0\sim3.1\times10^4\mathrm{V/m}$. 
These latter two cases run with Ti:Sa fundamental output, i.e, large $\hbar\Omega=1.55\mathrm{eV}$.
Table.~\ref{Table:pump_strength} lists a few typical $\gamma$ values. 
\begin{table}%The best place to locate the table environment is directly after its first reference in text
%\centering
\begin{tabular}{c c | c c c c}
 %& & \multicolumn{2}{c}{intrinsic}  \\
%\cline{2-5}
%\cmidrule(lr){3-4} %\cmidrule(lr){5-5}
%\cline{3-4} \cline{5-5}
% & & $\chi$ & $\nu$   \\
%\cline{2-4}
%\hline
%\cline{2-5}
%\cmidrule{2-4}  
& \multirow{2}{*}{$\gamma$} & \multicolumn{4}{c}{$\hbar\Omega/\mathrm{meV}$} \\
& & $\Delta/2=20$ & $\Delta=80$ & $2\Delta=80$ & Ti:Sa $1.55\times10^3$  \\ 
\cline{1-6} 
%\cmidrule{2-4} 
\multirow{2}{*}{$E_0/(10^5\mathrm{V/m})$} & $2.5$ & 0.083 & 0.021 & 0.0052 & 0.000026  \\
%\cline{2-2} 
 & $0.5$ & 0.017 & 0.0041 & 0.0010 & 0.0000052 %\\
\end{tabular}
\caption{Realistic estimation of dimensionless pump field strength. We assume an exchange gap $\Delta=55\mathrm{meV}$, two exemplary electric field strength $E_0=2.5,0.5\times10^5\mathrm{V/m}$, and Fermi velocity $v=0.4\times10^6\mathrm{m/s}$. Four different driving frequency are listed in relation to the gap size.}\label{Table:pump_strength}
\end{table}

%While the probe pulse is usually a femtosecond UV    laser pulse, the pump field can be a femtosecond pulse with frequency ranging from THz to visible light. Till now, time-resolved ARPES systems have achieved energy resolution of 10 to $10^2$ meV, time resolution of sub-fs to $10^2$fs\cite{Lv2019,Sobota2021}.   

We then estimate the tornado spiral arm width $k_\mathrm{arm}$. Based on Eq.~\eqref{eq:P3_main}, we have the simple phase relation
\begin{equation}
    2[d(k_\mathrm{arm})-d(0)]t/\hbar=2\pi,
\end{equation}
leading to 
\begin{equation}
    k_\mathrm{arm} = \frac{1}{\hbar v}\sqrt{(m+\frac{h}{2t})^2-m^2}.
\end{equation}
For instance, when $\Delta=70\mathrm{meV},v=0.3\times10^6\mathrm{m/s},t=0.5\mathrm{ps}$, we have $k_\mathrm{arm}=0.009\mathrm{\text{\AA}^{-1}}$.

We also estimate the strength of possible hexagonal warping effect in the dimensionless quantity 
\begin{equation}
    \lambda=c_6k_0^2/v
\end{equation}
with the characteristic momentum $k_0=\Delta/v$. Taking Bi$_2$Te$_3$ with $v=2.87\mathrm{eV\text{\AA}}, c_6=45.02\mathrm{eV\text{\AA}^3},\Delta=60\mathrm{meV}$ as an example, we have $\lambda=0.007\ll1$.

\section{Relaxation due to interaction effects}\label{SM:relaxation}
\begin{figure}[hbt]
\includegraphics[width=8.7cm]{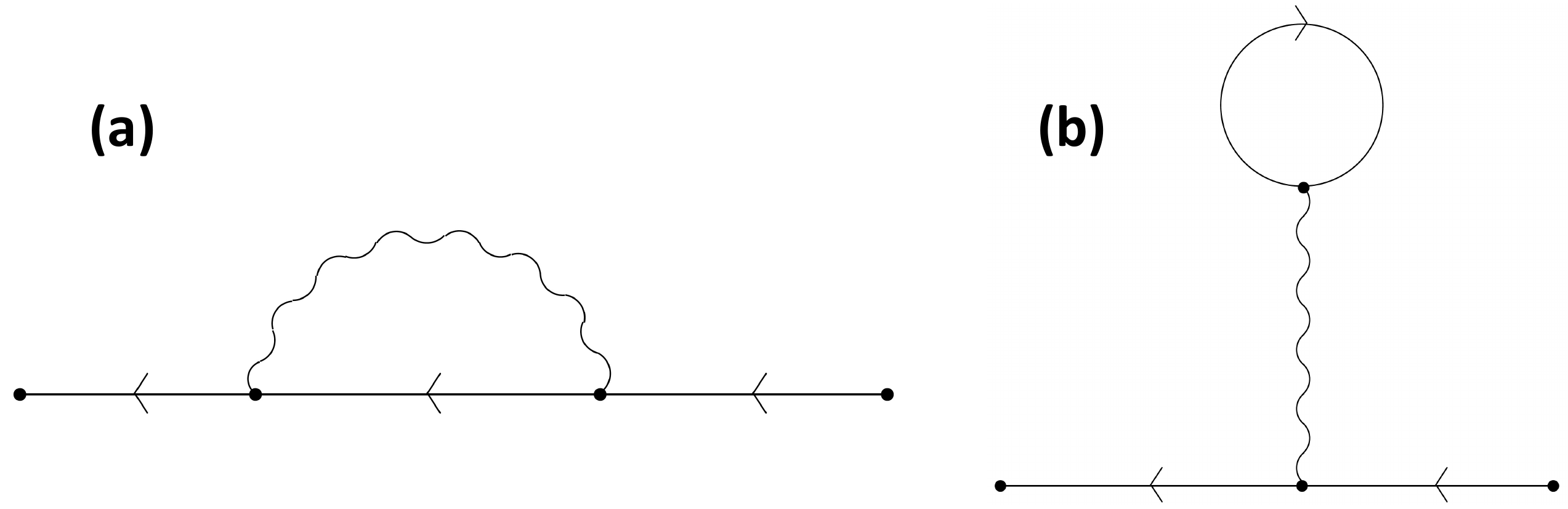}
\caption{Feynman diagrams of the lowest-order two possible interaction processes. Solid line denotes the electron propagator; wavy line denotes either the phonon propagator or the Coulomb interaction.
}\label{Fig:HFdiagrams}
\end{figure}

We briefly discuss the interaction effects from the viewpoint of relaxation and/or decoherence. For the solid-state system or more specifically the topological insulator surface state, there always exist multiple interaction channels, including the electron-lattice coupling, electron-electron interaction, disorder scattering, and random fluctuating electromagnetic field, etc.
%\subsubsection{Pertuabation theory}
Here, we exemplify the perturbative correction to the electronic Green's function with the electron-phonon interaction. The essential framework will remain the same for other interaction channels as well. We stick again to the Keldysh formalism. From the exact Dyson equation, we have 
\begin{equation}
\begin{split}
    G^\mathrm{r(a)} &= G_0^\mathrm{r(a)}(1+\Sigma^\mathrm{r(a)}G^\mathrm{r(a)})\\
    G^< &= (1+G^\mathrm{r}\Sigma^\mathrm{r}) G_0^<  (1+ \Sigma^\mathrm{a}G^\mathrm{a}) + G^\mathrm{r} \Sigma^< G^\mathrm{a}
\end{split}    
\end{equation}
where we always have the self-energies coming from the optical pump and the electron-phonon interaction $\Sigma=\Sigma_\mathrm{A}+\Sigma_\mathrm{I}$. The pure effect from optical pump $\Sigma_\mathrm{A}$ has been studied in detail in the main text. Compared to the notation in the main text, here we add the subscript $\mathrm{A}$ to distinguish it from $\Sigma_\mathrm{I}$.
Up to the low-order self-energy contributions, we have
\begin{equation}
\begin{split}
    G_{1\mathrm{I}}^\mathrm{r(a)} &= G_0^\mathrm{r(a)}\Sigma_\mathrm{I}^\mathrm{r(a)}G_0^\mathrm{r(a)}(1+\Sigma_\mathrm{A}^\mathrm{r(a)}G_0^\mathrm{r(a)})\\
    G_{1\mathrm{I}}^< &= G_0^\mathrm{r}\Sigma_\mathrm{I}^\mathrm{r} G_0^< (1+\Sigma_\mathrm{A}^\mathrm{a}G_0^\mathrm{a}) +  (1+\Sigma_\mathrm{A}^\mathrm{r}G_0^\mathrm{r}) G_0^< \Sigma_\mathrm{I}^\mathrm{a}G_0^\mathrm{a} + G_0^\mathrm{r}(1+\Sigma_\mathrm{A}^\mathrm{r}G_0^\mathrm{r}) \Sigma_\mathrm{I}^< G_0^\mathrm{a}(1+\Sigma_\mathrm{A}^\mathrm{a}G_0^\mathrm{a})- G_0^\mathrm{r}\Sigma_\mathrm{A}^\mathrm{r}G_0^\mathrm{r} \Sigma_\mathrm{I}^< G_0^\mathrm{a}\Sigma_\mathrm{A}^\mathrm{a}G_0^\mathrm{a}.
\end{split}    
\end{equation}
Each second term in the parentheses is bilinear in $\Sigma_\mathrm{A},\Sigma_\mathrm{I}$ and may cause combined effect. However, compared to the rest, these higher-order terms are of even smaller contribution in the weak coupling limit of our main interest. Note that the major effect of $\Sigma_\mathrm{A}$ is in the linear response and relevant experimental settings are estimated to be often deep in the weak-field regime as shown in Table.~\ref{Table:pump_strength}.
In the following, we hence focus on the leading interaction effect purely linear in $\Sigma_\mathrm{I}$
%In this regard, we should include the effect of $\Sigma_\mathrm{A}$ in the dressed electron Green's function  can 
\begin{equation}\label{eq:ep_dyson}
\begin{split}
    G_{1\mathrm{I}}^\mathrm{r(a)} &= G_0^\mathrm{r(a)}\Sigma_\mathrm{I}^\mathrm{r(a)}G_0^\mathrm{r(a)}\\
    G_{1\mathrm{I}}^< &= G_0^\mathrm{r}\Sigma_\mathrm{I}^\mathrm{r} G_0^<  +   G_0^< \Sigma_\mathrm{I}^\mathrm{a}G_0^\mathrm{a} + G_0^\mathrm{r} \Sigma_\mathrm{I}^< G_0^\mathrm{a}.
\end{split}    
\end{equation}
%For this part, due to the time-independence, it is convenient to work in the frequency domain. 
we take the simplest form of electron-phonon interaction and suppress polarization
\begin{equation}
    H_\mathrm{I}=\sum_{\bq\sigma} g_\bq c_{\bk+\bq\sigma}^\dag c_{\bk\sigma}(a_\bq+a_{-\bq}^\dag)+\mathrm{h.c.}
\end{equation}
where the phonon mode has the dispersion $\omega_\bq$. We henceforth denote the free phonon propagator 
\begin{equation}
    D_0(\bq,t,t')=-\ii\langle \mathrm{T}_\mathcal{C} Q_\bq(t) Q_{-\bq}(t')\rangle
\end{equation}
with $Q_\bq=a_\bq+a_{-\bq}^\dag$ in the Keldysh contour. According to Migdal's theorem, it suffices to drop vertex corrections for the dominating effects.
The leading diagrammatic contributing process to the self-energy thus possesses the paralell electron and phonon lines as shown in Fig.~\ref{Fig:HFdiagrams}(a). The Hartree diagram Fig.~\ref{Fig:HFdiagrams}(b) comes with a phonon propagator $D_0(\bq=0)$ at zero momentum connected to a fermion loop and thus affects the chemical potential only, which one can safely neglect in terms of the present discussion. Here we note that the electron-electron Coulomb interaction case is contributed by the same two diagrams in Fig.~\ref{Fig:HFdiagrams}. Since the main features remain essentially the same, we keep our focus on the electron-phonon case in the following.
Applying the various relations between the unrotated and rotated Keldysh Green's functions (Langreth rules)\cite{Stefanucci2015_SI}, which holds as well to self-energies, we have 
\begin{equation}\label{eq:ep_selfenergies}
\begin{split}
    \Sigma_{\mathrm{I}}^< &= G_0^<D_0^<\\
    \Sigma_{\mathrm{I}}^\mathrm{r} &= G_0^{\mathds{T}}D_0^{\mathds{T}}- G_0^<D_0^<=G_0^<D_0^\mathrm{r}+G_0^\mathrm{r}D_0^<+G_0^\mathrm{r}D_0^\mathrm{r}\\
    \Sigma_{\mathrm{I}}^\mathrm{a} &= G_0^{\mathds{T}}D_0^{\mathds{T}}- G_0^>D_0^>=G_0^<D_0^\mathrm{a}+G_0^\mathrm{a}D_0^<-G_0^\mathrm{a}D_0^\mathrm{a},
\end{split}    
\end{equation}
where we temporarily omit the interaction vertex for brevity.

Before proceeding, we need to specify the free propagators of the spinful electrons and the phonons
\begin{equation}
\begin{split}
    G_0^<(\bk,\omega) &= 2\pi\ii \sum_a |\bk a\rangle \langle \bk a|f_{\bk a}\delta(\omega-\varepsilon_{\bk a}) \\
    G_0^\mathrm{r(a)}(\bk,\omega) &=\sum_{a}|\bk a\rangle \langle \bk a| \frac{1}{\omega - \varepsilon_{\bk a} \pm \ii\eta}\\
    D_0^<(\bk,\omega) &= -2\pi\ii [(n_{\bq}+1)\delta(\omega+\omega_\bq)+n_{\bq}\delta(\omega-\omega_\bq)] \\
    D_0^\mathrm{r(a)}(\bk,\omega) &= \frac{1}{\omega - \omega_{\bq} \pm \ii\eta} - \frac{1}{\omega + \omega_{\bq} \pm \ii\eta},
\end{split}    
\end{equation}
where $f_{\bk a}$ is the Fermi distribution for the electron in band basis $a=\pm$ and $n_\bq$ is the phonon distribution.
We now concretely evaluate Eq.~\eqref{eq:ep_selfenergies} with the shorthand $\bk'=\bk-\bq$
\begin{equation}\label{eq:ep_selfenergies1}
\begin{split}
    %\Sigma_{\mathrm{I}}^<(\bk,t,t') &= \ii \sum_\bq |g_\bq|^2 G_0^<(\bk-\bq,t-t')D_0^<(\bq,t-t')\\
    \Sigma_{\mathrm{I}}^<(\bk,\omega) &= \ii \int\frac{\dd\varepsilon}{2\pi}\sum_\bq |g_\bq|^2 G_0^<(\bk',\omega-\varepsilon)D_0^<(\bq,\varepsilon)\\
    &= 2\pi\ii \sum_{\bq a} |g_\bq|^2  |\bk' a\rangle \langle \bk' a|f_{\bk' a}[(n_{\bq}+1)\delta(\omega-\varepsilon_{\bk'a}+\omega_\bq)+n_{\bq}\delta(\omega-\varepsilon_{\bk'a}-\omega_\bq)]\\
    \Sigma_{\mathrm{I}}^\mathrm{r}(\bk,\omega) &= \ii \int\frac{\dd\varepsilon}{2\pi} \sum_\bq |g_\bq|^2 [G_0^<(\bk',\omega-\varepsilon)D_0^\mathrm{r}(\bq,\varepsilon)+G_0^\mathrm{r}(\bk',\omega-\varepsilon)D_0^<(\bq,\varepsilon)+G_0^\mathrm{r}(\bk',\omega-\varepsilon)D_0^\mathrm{r}(\bq,\varepsilon)]\\
    &= \sum_{\bq a} |g_\bq|^2  |\bk' a\rangle \langle \bk' a|\left[ \frac{n_\bq+1-f_{\bk'a}}{\omega-\varepsilon_{\bk'a}-\omega_\bq+\ii\eta} + \frac{n_\bq+f_{\bk'a}}{\omega-\varepsilon_{\bk'a}+\omega_\bq+\ii\eta} \right]\\
    \Sigma_{\mathrm{I}}^\mathrm{a}(\bk,\omega) &= {\Sigma_{\mathrm{I}}^\mathrm{r}}^\dag(\bk,\omega) .
\end{split}    
\end{equation}
These expression can readily be used to calculate the lowest order correction Eq.~\eqref{eq:ep_dyson} in the electron Green's function. 
%For instance, for the correction to retarded Green's function we have
% \begin{equation}\label{eq:G1r}
% \begin{split}
%     G_{1\mathrm{I}}^\mathrm{r}(\bk,t_1,t_2) &= \int\dd t_3\dd t_4\sum_{\bk_3\bk_4}G_0^\mathrm{r}(\bk,t_1;\bk_3,t_3)\Sigma_\mathrm{I}^\mathrm{r}(\bk_3,t_3;\bk_4,t_4)G_0^\mathrm{r}(\bk_4,t_4;\bk,t_2)\\
%     &= \int\frac{\dd\varepsilon}{2\pi} \ee^{-\ii\varepsilon(t_1-t_2)} G_0^\mathrm{r}(\bk,\varepsilon)\Sigma_\mathrm{I}^\mathrm{r}(\bk,\varepsilon)G_0^\mathrm{r}(\bk,\varepsilon)\\
%     &= \int\frac{\dd\varepsilon}{2\pi} \sum_{\bq abc}\ee^{-\ii\varepsilon(t_1-t_2)}  \frac{|g_\bq|^2   |\bk a\rangle \langle \bk c| \braket{\bk a|\bk'b}\braket{\bk'b|\bk c}}{(\varepsilon-\varepsilon_{\bk a}+\ii\eta)(\varepsilon-\varepsilon_{\bk c}+\ii\eta)} \left[ \frac{n_\bq+1-f_{\bk'b}}{\varepsilon-\varepsilon_{\bk'b}-\omega_\bq+\ii\eta} + \frac{n_\bq+f_{\bk'b}}{\varepsilon-\varepsilon_{\bk'b}+\omega_\bq+\ii\eta} \right].
% \end{split}    
% \end{equation}
For instance, we can look at the retarded  $\Sigma_\mathrm{I}^\mathrm{r}$ in Eq.~\eqref{eq:ep_selfenergies1}. The process of absorption and emission of one phonon of momentum $\bq$ manifests in the energy factors. %Also note from the spin matrix structure that an electron after emission or absorption of a phonon in general changes its spin state and can have finite projection onto the particular final state. 
Such inelastic scattering processes gives rise to the relaxation of the original spin-orbit coupled electronic state and hence the decoherence or a finite lifetime.

This formalism displays how one can take into account the interaction effects and consider the corresponding interaction-induced correction to the various single-particle electronic Green's functions from the Keldysh-contour $G(\bk,t_1,t_2)$ with spin degree of freedom, which are relevant to what SARPES measures experimentally.
The characteristic relaxation times can be estimated from the quasiparticle lifetime embedded in the retarded self-energy. Because of the general matrix relation $(G^\mathrm{r})^{-1}=(G_0^\mathrm{r})^{-1}-\Sigma_\mathrm{I}^\mathrm{r}$, we switch the representation of $\Sigma_\mathrm{I}^\mathrm{r}$ to the eigenbasis $|\bk a\rangle$ that diagonalizes $H_0=\bd(\bk)\cdot\bsigma$ and hence $G_0^\mathrm{r}$. We therefore denote the unit vector $\hat{\bd}$ in the diagonal basis and another unit vector $\hat{\bd}^\perp$ normal to $\hat{\bd}$ for the off-diagonal entries. We can make the following identification of the relaxation time scales respectively for the band-diagonal and band-offdiagonal contributions
\begin{equation}\label{eq:retardedT1T2}
\begin{split}
    T_1^{-1}\sim-2\mathrm{Im}\mathrm{Tr}[\hat{\bd}\cdot\bsigma\Sigma_\mathrm{I}^\mathrm{r}] &= 2\pi \sum_{\bq a} K_{\bq a} \left[ (n_\bq+1-f_{\bk-\bq, a})\delta(\omega-\varepsilon_{\bk-\bq, a}-\omega_\bq) + (n_\bq+f_{\bk-\bq, a})\delta(\omega-\varepsilon_{\bk-\bq, a}+\omega_\bq) \right]\\
    T_2^{-1}\sim-2\mathrm{Im}\mathrm{Tr}[\hat{\bd}^\perp\cdot\bsigma\Sigma_\mathrm{I}^\mathrm{r}] &= 2\pi \sum_{\bq a}  K^\perp_{\bq a} \left[ (n_\bq+1-f_{\bk-\bq, a})\delta(\omega-\varepsilon_{\bk-\bq, a}-\omega_\bq) + (n_\bq+f_{\bk-\bq, a})\delta(\omega-\varepsilon_{\bk-\bq, a}+\omega_\bq) \right],
\end{split}    
\end{equation}
where we denote $K_{\bq a}=|g_\bq|^2\mathrm{Tr}[\hat{\bd}\cdot\bsigma|\bk-\bq, a\rangle \langle \bk-\bq, a|]$ and $K^\perp_{\bq a}=|g_\bq|^2\mathrm{Tr}[\hat{\bd}^\perp\cdot\bsigma|\bk-\bq, a\rangle \langle \bk-\bq, a|]$. Note that these two scales are momentum- and frequency-dependent as per the Green's function relation, where the physically relevant frequency is typically given by the band gap. Here we use the notation that $T_1$ is mainly for band energy relaxation and $T_2$ is for the interband decoherence time.
Theoretically, as shown in Eq.~\eqref{eq:retardedT1T2}, because of the common origin from electron-phonon or electron-electron interaction, it is natural to expect that $T_1\sim T_2$ holds in general. Indeed, often a comparison in the range from $T_2\approx0.5T_1$ to $T_2\lesssim2T_1$ is observed in electronic spin experiments\cite{Bar_Gill_2013_SI,Sigillito_2015_SI}.
For topological insulator surface state, usually $T_1$ is more accessible and estimated from spin-resolved spectroscopies to be at the order of $4\textrm{-}15\mathrm{ps}$\cite{Hosur2011_SI,Cacho2015_SI,Iyer2018_SI}. This thus guarantees a coherence time $T_2$ at the same order, which is sufficient to observe the fine tornado patterns of our main interest, since these patterns rely on the interband quantum coherence. %As a matter of fact, lower temperature of the bulk system will always help maintain the coherence longer as it will enter through the $\gamma_{\alpha\beta}$ in Eq.~\eqref{eq:lindbladian}. 
Also interestingly, from a quantum Boltzmann equation approach the topological insulator surface state is shown to have both the out-of-plane and in-plane spin relaxation times locked to twice the momentum relaxation time\cite{liu_2013_SI}. This can be regarded as an idealized yet still supporting evidence towards realistic detection of the physical information in the spin channel.

\end{document}